\documentclass[prb,twocolumn,showpacs,10pt]{revtex4-1}
\usepackage{textcomp}
\usepackage{amsmath}
\usepackage{amsfonts}
\usepackage{calc}
\usepackage{mathrsfs}
\usepackage{amssymb}
\usepackage{amsmath}
\usepackage{array}
\usepackage{color}
\usepackage{bm}
\usepackage{graphicx}
\usepackage{hyperref}

\newcommand{\lam}{\lambda}
\newcommand{\sxx}{{\rm Re}\ \sigma_{xx}}
\newcommand{\sxy}{{\rm Im}\ \sigma_{xy}}
\renewcommand{\sp}{{\rm Re}\ \sigma_+}
\newcommand{\sm}{{\rm Re}\ \sigma_-}
\newcommand{\spm}{{\rm Re}\ \sigma_\pm}

\begin{document}
\title{Magneto-optical conductivity in graphene including electron-phonon coupling} 
\author{Adam Pound$^1$} 
\author{J.P. Carbotte$^{2,3}$}
\author{E.J. Nicol$^4$}
\affiliation{$^1$School of Mathematics, University of Southampton, Southampton, United Kingdom, SO17 1BJ}
\affiliation{$^2$Department of Physics and Astronomy, McMaster University, Hamilton, Ontario, Canada, L8S 4M1}
\affiliation{$^3$The Canadian Institute for Advanced Research, Toronto, Ontario, Canada, M5G 1Z8}
\affiliation{$^4$Department of Physics, University of Guelph, Guelph, Ontario, Canada, N1G 2W1}
\pacs{78.67.Wj, 71.70.Di, 63.22.Rc, 73.22.Pr}
\date{\today}

\begin{abstract}
We show how coupling to an Einstein phonon $\omega_E$ affects the absorption peaks seen in the optical conductivity of graphene under a magnetic field $B$. The energies and widths of the various lines are shifted, and additional peaks arise in the spectrum. Some of these peaks are Holstein sidebands, resulting from the transfer of spectral weight in each Landau level (LL) into phonon-assisted peaks in the spectral function. Other additional absorption peaks result from transitions involving split LLs, which occur when a LL falls sufficiently close to a peak in the self-energy. We establish the selection rules for the additional transitions and characterize the additional absorption peaks. For finite chemical potential, spectral weight is asymmetrically distributed about the Dirac point; we discuss how this causes an asymmetry in the transitions due to left- and right-handed circularly polarized light and therefore oscillatory behavior in the imaginary part of the off-diagonal Hall conductivity. We also find that the semiclassical cyclotron resonance region is renormalized by an effective-mass factor but is not directly affected by the additional transitions. Last, we discuss how the additional transitions can manifest in broadened, rather than split, absorption peaks due to large scattering rates seen in experiment.
\end{abstract}
\maketitle

\section{Introduction}
\begin{figure*}[tb]
\begin{center}
\includegraphics[width=0.9\textwidth]{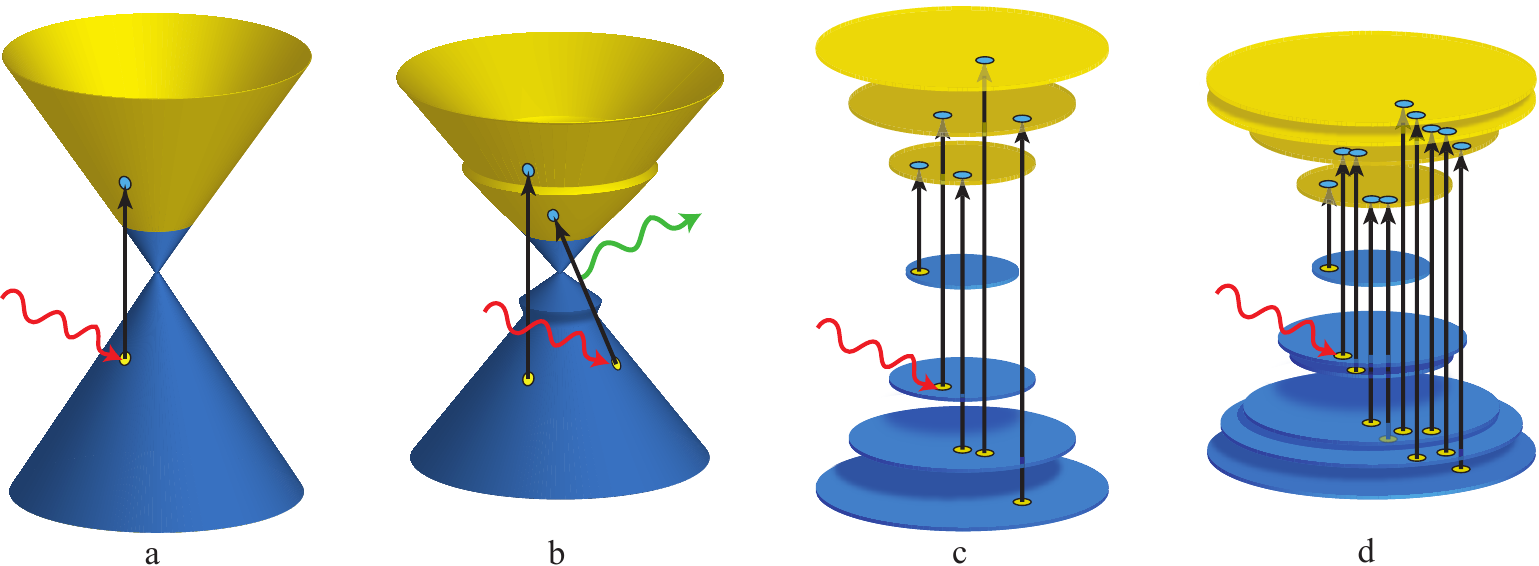}
\end{center}
\caption{(Color online) Schematic depiction of energy bands or levels in graphene and the possible optical transitions between them. In all cases, blue indicates a filled state and yellow indicates an empty one. Each transition is excited by an incoming photon (indicated by a red arrow), but we show only one photon in each diagram to prevent overcrowding. (a) Bare bands forming two cones meeting at the Dirac point, with a photon exciting an interband transition from the valence band to an unoccupied state in the conduction band. (b) Bands dressed by coupling to a phonon, with widened cones and kinks at plus or minus the phonon frequency (relative to the Fermi energy). The emission of a phonon (indicated by the green arrow) allows additional transitions which can occur at lower energies and can alter the quasiparticle momentum. (c) Bare bands condensed into discrete Landau levels when under an applied magnetic field, with an intraband transition from the level just below the Fermi energy to the one just above, and interband transitions from levels labeled by negative integers $-n$ ($n$ positive or zero) to levels $n\pm1$. (d) Landau levels dressed by coupling to a phonon, introducing split levels and phonon-assisted peaks, allowing additional transitions between them. \label{cones}}
\end{figure*}

At low energies, the charge-carrier dynamics in graphene are governed by the Dirac equation for massless fermions, leading to many unusual properties that are now well documented in several review articles.\cite{Geim:07, Castro-Neto:09, Gusynin:07c, Abergel:10, Orlita:10} The charge carriers in a single layer can be described by chiral quasiparticles with a linear energy dispersion.
With the two atoms per unit cell, there are two sets of conic bands at two $K$ points in the corresponding Brillouin zone. In each of the two sets, the apexes of the two cones coincide at a point called the Dirac point, with the upper, upright cone forming the conduction band and the lower, inverted cone forming the valence band, as depicted schematically in Fig.~\ref{cones}a. The existence of the Dirac point and cones has been verified in many experiments.\cite{Geim:07, Castro-Neto:09, Gusynin:07c, Abergel:10, Orlita:10} In particular, angle-resolved photoemission experiments (ARPES)\cite{Bostwick:07, Zhou:08, Bianchi:10, Bostwick:10} have directly measured the electronic dispersion curves, and scanning tunneling spectroscopy (STS) experiments\cite{Zhang:08, Li-G:09, Miller:09, Brar:10} have measured the corresponding\cite{Nicol:09} linear-in-energy density of states (DOS).

The double-Dirac-cone band structure in graphene leads directly to interband optical transitions,\cite{Ando:02, Gusynin:06b, Gusynin:09, Falkovsky:07, Falkovsky:08, Stauber:08b} as depicted by the vertical arrow in Fig.~\ref{cones}a. In this process a quasiparticle is promoted from an occupied valence state below the Fermi energy (blue region) to an empty conduction state above (yellow region). With doping away from charge neutraliy, the chemical potential $\mu$ becomes finite (as shown in Fig.~\ref{cones}a), and the transitions then require photons of energy $\Omega\geq 2|\mu|$, due to Pauli exclusion. At energies above this threshold, the conductivity displays a constant universal background value equal to $\sigma_0=\frac{e^2}{4\hbar}$, where $e$ is the electron charge. In addition to the interband transitions, there are the usual intraband ones involving only the conduction band, which provide the usual Drude peak centered around a photon energy $\Omega=0$. Away from the charge neutrality point ($\mu=0$), one can expect $2|\mu|$ to be much larger than the scattering rate associated with the Drude absorption, leading to a region of near-zero absorption between the Drude peak and the onset of the universal background. This predicted behavior has been observed in optical measurements by several groups.\cite{Li:08, Wang:08, Mak:08, Nair:08, Orlita:10} Associated work on bilayer graphene has also been performed.\cite{Nair:08, Nicol:08, Li-Z:09, Kuzmenko:09}

While experiments\cite{Bostwick:07, Zhou:08, Bianchi:10, Bostwick:10, Li-G:09, Miller:09} have largely confirmed the bare-band picture of graphene, signatures of many-body corrections have also been seen. Specifically, kinks appear in the dressed dispersion curves measured by ARPES,\cite{Bostwick:07, Zhou:08, Bianchi:10, Bostwick:10} which are interpreted as the result of coupling to a phonon of energy $\omega_E$, as shown schematically in Fig.~\ref{cones}b. Corresponding structures have been seen in STS measurements as well.\cite{Li-G:09, Miller:09, Brar:10, Pound:11a} This phonon structure, and other phonon effects, are particularly prominent in graphene, unlike in conventional metals, because graphene's electronic DOS varies with energy on the scale of the phonon frequency.\cite{Mitrovic:83a, Mitrovic:83b, Nicol:09} There also exists striking evidence for plasmaron signatures coming from electron-electron interactions: ARPES spectra\cite{Bostwick:10} have found that the Dirac point where the two bands meet is split in two, with an extended plasmaron region between them. The magnitude of such effects, however, depends on the substrate dielectric constant $\epsilon$, since the screened Coulomb potential is inversely proportional to $\epsilon$. For example, for graphene on H/SiC, $\epsilon=2$, while on SrTiO$_3$, $\epsilon$ can be varied by more than an order of magnitude and reach 5000 when there is a change from room to liquid He temperature.\cite{Couto:11} Thus, in principle, electron-electron effects can be switched off by a judicious choice of substrate, while phonon effects remain.

These many-body effects also alter the picture of the optical conductivity described above. In particular, experiments have not observed precisely the near-zero behavior in the region between the Drude peak at $\Omega=0$ and the onset of interband absorption at $\Omega=2\mu$.\cite{Gusynin:09} Instead of dropping to zero, the absorption never falls below about $1/3$ of its universal background value $\sigma_0$.\cite{Li:08} This can be understood in terms of electron-phonon renormalization effects. In conventional metal physics, it is well known that besides a Drude peak there exist phonon-assisted sidebands.\cite{Mori:08, Carbotte:10} These sidebands arise due to Holstein processes, in which a phonon is created along with a particle-hole excitation, as shown in Fig.~\ref{cones}b. Such sidebands have been exploited to get detailed information about the electron-phonon spectral density.\cite{Farnsworth:74, Carbotte:90} More recently, they have also yielded valuable information on the low-energy excitations that couple to charge in the high-$T_c$ oxides, interpreted as spin fluctuations.\cite{Hwang:06,Hwang:07} While the Holstein processes associated with both intra and interband transitions in graphene are not expected to be large,\cite{Bostwick:07, Li-G:09, Nicol:09, Carbotte:10, Park:07, Park:08, Park:09} they are predicted\cite{Carbotte:10} to contribute a significant part, though not all, of the absorption seen in the Pauli-blocked region of the conductivity spectrum. Impurities may contribute as well,\cite{Peres:08} as might electron-electron processes,\cite{Grushin:09} although that has yet to be established. If the electron-electron interaction does contribute, the magnitude could be manipulated by performing experiments on a variety of substrates, allowing one to isolate the electron-phonon contribution.

When a magnetic field $B$ is applied perpendicularly to the graphene plane, the Dirac cones are transformed into discrete Landau levels at energies $M_n$ (relative to the Dirac point), as illustrated in Fig.~\ref{cones}c. (Note that the disks are schematic; technically each level is made up of concentric rings.)\cite{Roldan:10, Pound:11b} Because of the Dirac nature of the quasiparticles, the levels have the relativistic form $M_n\propto{\rm sgn}(n)\sqrt{|n|B}$ for integer $n$. In this case a photon can induce a transition between two LLs.\cite{Gusynin:07a} Beyond conserving the total energy, the intraband transitions must obey the selection rules $n\to n+1$ (where $n$ is any integer), and the interband transitions must obey $-n\to n+1$ or $-(n+1)\to n$ (where here $n$ is non-negative), where in all cases the initial state must be occupied and the final state unoccupied;\cite{Gusynin:07a, Gusynin:07b, Gusynin:07d, Bychkov:08, Gusynin:09} see again Fig.~\ref{cones}c. This leads to a well-defined sequence of absorption peaks in the conductivity, some of which have been
seen in experiment.\cite{Sadowski:06, Sadowski:07, Jiang:07, Henriksen:10, Deacon:07}
There are also recent experimental results 
on bilayer graphene.\cite{Li-Z:09,Henriksen:08}

In this paper we study how those absorption peaks are modified in a simple model of constant coupling to a single Einstein phonon of frequency $\omega_E$. Our previous work on the DOS\cite{Pound:11a, Pound:11b} showed that in the presence of such coupling, energy levels are not only shifted and broadened, but two additional types of peaks arise in the spectrum: phonon-assisted peaks at energies $E_n\pm\omega_E$ corresponding to peaks in the self-energy, where $E_n$ is the energy of the $n$th LL relative to the Fermi energy; and split LLs that occur when the renormalized levels lie sufficiently close to a peak in the self-energy. These additional peaks allow additional transitions not seen in the bare spectrum, as shown schematically in Fig.~\ref{cones}d. Hence, the optical conductivity will have additional absorption peaks and sidebands corresponding to these additional transitions.

We begin in Sec.~\ref{formalism} by reviewing the formalism and pertinent results from our previous work\cite{Pound:11a,Pound:11b} and summarizing the nature of the renormalization effects that will be discussed in the subsequent sections. Section~\ref{spectral_functions} follows that by analysing the dressed form of the spectral functions associated with individual LLs, which allows us to understand the form of the conductivity. Here we introduce the notation and definitions for peaks and transitions. To keep these first two sections streamlined, derivations of selection rules are relegated to appendices. In Sec.~\ref{general behavior}, we discuss the general features of the dressed conductivity, including the field-dependence of additional absorption peaks and their optical weights. In Sec.~\ref{chemical potential}, we explore the effects of varying the chemical potential. Section~\ref{circular} examines the conductivity for circularly polarized light. Section~\ref{cyclotron} shows how the semiclassical cyclotron resonance is renormalized. In Sec.~\ref{experiment}, we discuss the effects of broadening and the bearing of our results on recent experiments. We summarize and conclude in Sec.~\ref{summary}.


\section{Formalism}\label{formalism}

In the absence of a magnetic field, the linear dispersion rising out of the Dirac point, in the continuum limit of a simple nearest-neighbour tight-binding Hamiltonian, is given by $\epsilon_k=\hbar v_F k$,
where $v_F$ is the Fermi velocity and $k$ is the magnitude of momentum measured relative to the Dirac point. Here we will always (except in Sec.~\ref{experiment}) use the typical value $v_F=10^6$m/s. This dispersion leads to the linear DOS $N^0(\omega)=N_0|\omega+\mu_0|$, where $\mu_0$ is the non-interacting chemical potential, $N_0\equiv\frac{2}{\pi\hbar^2v_F^2}$, and the superscript zero indicates that this is the bare DOS. In the presence of a magnetic field $B$, the energy dispersion condenses into discrete
Landau levels $M_n={\rm sgn}(n) v_F\sqrt{2|n|eB\hbar/c}$ (in Gaussian units), where $n$ is any integer and $v_F$ is the Fermi velocity. The DOS then becomes a sum over level index $n$:
\begin{align}
N^0(\omega)=\frac{1}{2}N_0 M_1^2\theta(W_C-|\omega+\mu_0|)\!\!\sum_{n=-\infty}^\infty
\!\!\delta(\omega+\mu_0-M_n),\label{DOSbare}
\end{align}
which consists of a line for each Landau level, at energies $E_n=M_n-\mu_0$ relative to the Fermi energy. Here $W_C$ is a high-energy cutoff on the linear dispersion approximation. Throughout this paper, we use the cutoff $W_C=\sqrt{\pi\sqrt 3}t$, where $t$ is the nearest-neighbor hopping parameter, which ensures that the number of states in the Dirac cones equals the number in the first Brillouin zone. Specifically, we use $W_C=7$eV, corresponding to the typical value $t\simeq 3$eV.

When the electron-phonon coupling is taken into account, this formula is generalized to\cite{Sharapov:04}
\begin{align}
N(\omega)=\frac{1}{2}N_0M_1^2\theta(W_C-|\tilde\omega|)\!\!\sum_{n=-\infty}^\infty
A_n(\omega),\label{DOS}
\end{align}
where $\tilde\omega\equiv\omega-\Sigma_1(\omega)+\mu$ and 
$A_n(\omega)$ is the spectral function for the $n$th Landau level, 
\begin{align}
A_n(\omega) = \frac{1}{\pi}\frac{\Gamma-\Sigma_2(\omega)}{\left[\omega-\Sigma_1(\omega)+\mu-M_n\right]^2+\left[\Gamma-\Sigma_2(\omega)\right]^2},\label{dosA}
\end{align}
which is straightforwardly obtained from the full spectral function $A(\vec k,\omega)$ given in Ref.~\cite{Sharapov:04}. This form includes both many-body renormalizations, manifested in the self-energy $\Sigma=\Sigma_1+i\Sigma_2$, and a residual scattering rate $\Gamma$. The fully interacting chemical potential $\mu$ that enters here is evaluated from $\mu=\mu_0+\Sigma_1(0)$. For simplicity, we take $\Gamma$ to be constant; in a more comprehensive calculation, it would depend on both $\omega$ and $n$. One can easily see that in the limit of no interactions and $\Gamma\to 0$, Eqs.~\eqref{DOS} and \eqref{dosA} reduce to the bare case of $A_n(\omega)=\delta(\omega+\mu_0-M_n)$ and Eq.~\eqref{DOSbare}.

Note that Eq.~\eqref{dosA} is written
for a $k$-independent self-energy, as would be the case for a model of
the electron-phonon interaction where all $k$-information has been subsumed
into the frequency-dependent electron-phonon spectral function 
$\alpha^2F(\nu)$. This then allows the self-consistent self-energy to be
calculated at zero temperature by\cite{Dogan:03, Nicol:09, Carbotte:10}
\begin{align}
\Sigma(\omega) &= \frac{1}{W_C}\int_0^\infty d\nu\alpha^2F(\nu)\int^\infty_{-\infty}d\omega'\frac{N(\omega')}{N_0}\nonumber\\
&\quad\times\left[\frac{\theta(\omega')}{\omega-\nu-\omega'+i0^+}+\frac{\theta(-\omega')}{\omega+\nu-\omega'+i0^+}\right],\label{Sigma}
\end{align}
which is solved iteratively together with Eq.~\eqref{DOS}. 
To evaluate the self-energy, we use a simple model for the phonon spectrum suggested by
Park et al.\cite{Park:07} Using full first principle
calculations of the electron-phonon interaction in graphene, these authors
showed that the self-energy $\Sigma(\omega)$ can be well approximated by an Einstein phonon spectrum with
$\alpha^2F(\nu)=A\delta(\nu-\omega_E)$, where $\omega_E=200$meV and $A$ is the electron-phonon coupling strength. 
Here we assume that it remains a good approximation in the case of a finite field. We shall choose values of $A$ that give realistic values of the electron-phonon effective mass renormalization parameter $\lam\equiv-\frac{d\Sigma_1}{d\omega}(0)$, which is generally found to be $\sim0.1$. Even if the model proves less accurate in this case, due to matrix elements in the self-energy, for example, studying the coupling to a single phonon provides a simple means of understanding and characterizing the effects of coupling to any distribution of them. And so long as the full self-energy can still be well approximated by a coupling to a small number of phonon frequencies, our results will still apply for each of them; the appropriate values of $\omega_E$ and $A$ for each of the phonons must simply be determined by experiment. Additionally, the model is generic enough to apply to coupling to phonons associated with the substrate. The model is also indifferent to whether the phonons are IR-active. The utility of the model shows itself immediately in allowing us to easily find the form of the self-energy: performing one iteration of Eq.~(\ref{Sigma}), starting with the non-interacting DOS of Eq.~(\ref{dosA}), we find
\begin{align}
\Sigma(\omega) &= \frac{AM_1^2}{2W_C}\sum_{n=-n_{\rm max}}^{n_{\rm max}}\left[\frac{\theta(M_n-\mu_0)}{\omega-\omega_E-M_n+\mu_0+i0^+}\right.\nonumber\\
&\quad +\left.\frac{\theta(-M_n+\mu_0)}{\omega+\omega_E-M_n+\mu_0+i0^+}\right],\label{Sigmaiter}
\end{align}
where $n_{\rm max}$ is the largest integer smaller than $\frac{W^2_C}{M_1^2}$. The imaginary part of this self-energy has $\delta$-function peaks at $\omega=P_n\equiv \pm\omega_E+M_n-\mu_0$, and the real part has corresponding singularities at the same energies. In other words, the self-energy contains peaks corresponding to each of the LLs, but shifted by $\pm\omega_E$. Because of the Heaviside functions, these peaks always occur outside the interval $(-\omega_E,\omega_E)$; that is, the sign in front of $\omega_E$ in $P_n$ is always such that $|P_n|\geq\omega_E$. One can easily show\cite{Pound:11b} analytically that introducing a broadening $\Gamma$ produces two effects: logarithmic divergences at $\pm\omega_E$ that grow with $\Gamma$; and a slow change with varying $\omega$, due to the addition of tails of broadened peaks. Between peaks, this leads to the real part of the self-energy varying approximately as $\Sigma_1(\omega)=\Sigma_1(0)-\lam\omega$. All of this assumes a single iteration of Eqs.~\eqref{DOS} and \eqref{Sigma}, and throughout this paper, for simplicity we stop after one iteration. As discussed in our article on the DOS,\cite{Pound:11b} if the equations were iterated to convergence, the energies $P_n$ would be shifted by $\pm\omega_E$ relative to the dressed, rather than the bare, LLs, and multiphonon processes would give rise to additional peaks at energies shifted by multiples of $\omega_E$. For energies of magnitude below $2\omega_E$, the net effect is simply to increase the value of $\lam$, which is indistinguishable from increasing $A$. For energies above $2\omega_E$, the level structure will be obscured by the preponderance of peaks, unless the levels are very widely spaced relative to $\omega_E$. Since peaks at such high energies are not well resolved in experiment, this is not a significant limitation.

As we shall see in detail in the following section, incorporating the self-energy into the spectral functions of Eq.~\eqref{dosA} shifts the bare peak and introduces additional ones in a simple way. And determining these effects on the spectral functions is sufficient to determine the effects of electron-phonon coupling on the magneto-optical conductivity. Adapting the results of Gusynin et al.\cite{Gusynin:07a} to our notation, we find that the real part of the longitudinal optical conductivity, $\sxx$, and the imaginary part of the transverse conductivity, $\sxy$, at zero temperature are given by the following sums:
\begin{align}
\sxx(\Omega) &= \sigma_0\frac{M_1^2}{\Omega}\sum_{n=0}^\infty\int_0^\Omega d\omega\left[\psi_{n,n+1}(\omega,\omega-\Omega)\right.\nonumber\\
&\quad +\left.\psi_{n,n+1}(\omega-\Omega,\omega)\right],\label{sigma_xx}\\
\sxy(\Omega) &= \sigma_0\frac{M_1^2}{\Omega}\sum_{n=0}^\infty\int_0^\Omega d\omega\left[\psi_{n,n+1}(\omega,\omega-\Omega)\right.\nonumber\\
&\quad -\left.\psi_{n,n+1}(\omega-\Omega,\omega)\right],\label{sigma_xy}
\end{align}
where
\begin{align}
\psi_{n,m}(\omega,\omega') &= A_n(\omega)A_m(\omega')+A_{-n}(\omega)A_{-m}(\omega')\nonumber\\
& \quad +A_n(\omega)A_{-m}(\omega')+A_{-n}(\omega)A_m(\omega').\label{psi}
\end{align}
We shall denote the summands in these expressions by $\sigma^{(n)}_{ij}$, such that $\sigma_{ij}=\sum_{n=0}^\infty\sigma^{(n)}_{ij}$. Note that while the LLs and spectral functions are indexed by any integer, the conductivity is decomposed into pieces indexed by non-negative integers only.  Each piece of the conductivity, $\sigma^{(n)}_{ij}$, is determined by the overlap of pairs of spectral functions, corresponding to a transition between an initial and final level. In both the bare and dressed cases, four combinations of spectral functions contribute absorption peaks: $A_{n+1}(\omega)A_{n}(\omega-\Omega)$ and $A_{-n}(\omega)A_{-(n+1)}(\omega-\Omega)$, corresponding to intraband transitions $n\to n+1$ and $-(n+1)\to -n$, respectively; and $A_{n+1}(\omega)A_{-n}(\omega-\Omega)$ and $A_{n}(\omega)A_{-(n+1)}(\omega-\Omega)$, corresponding to the interband transitions $-n\to n+1$ and $-(n+1)\to n$, respectively. (Note again that $n$ here is positive or zero.) More details are provided in Appendix~\ref{selection_rules}, and the other combinations of spectral functions appearing in Eqs.~\eqref{sigma_xx} and \eqref{sigma_xy} are discussed in Appendix~\ref{inverted_transitions}.

In the bare case, each $A_n$ is a simple Lorentzian centered at $E_n=M_n-\mu_0$, from which it follows that each $\sigma^{(n)}_{ij}$ is simply a sum of Lorentzians. For $\Gamma=0$, Eq.~\eqref{sigma_xx} is easily evaluated to find a simple set of lines:
\begin{align}
&\sxx(\Omega) = \nonumber\\
&\quad \frac{\sigma_0 M_1^2}{\Omega}\sum_{n=0}^\infty\big[\delta\left(E_{n+1}-E_{n}-\Omega\right)\theta\left(E_{n+1}\right)\theta\left(-E_{n}\right)\nonumber\\
&\quad +\delta\left(E_{-n}-E_{-(n+1)}-\Omega\right)\theta\left(E_{-n}\right)\theta\left(-E_{-(n+1)}\right)\nonumber\\
&\quad +\delta\left(E_{n+1}-E_{-n}-\Omega\right)\theta\left(E_{n+1}\right)\theta\left(-E_{-n}\right)\nonumber\\
&\quad +\delta\left(E_{n}-E_{-(n+1)}-\Omega\right)\theta\left(E_{n}\right)\theta\left(-E_{-(n+1)}\right)\big],\label{sigmaxx_bare}
\end{align}
which follows from Eq.~\eqref{sigma_general} and was earlier derived by Gusynin et al.\cite{Gusynin:07a} The first two $\delta$-functions come from the intraband transitions; the latter two, from the interband.  The Heaviside functions state that the final state must be above the Fermi energy and the initial state must be below it. Note that because $M_{-n}=-M_n$, in the bare case we have $E_{n+1}-E_{-n}=E_{n}-E_{-(n+1)}$, making the interband transitions degenerate. This means each $\sxx^{(n)}$ has at most two peaks, one from an intraband transition and one from interband transitions.

\begin{figure}[tb]
\begin{center}
\includegraphics[height=0.6\textheight]{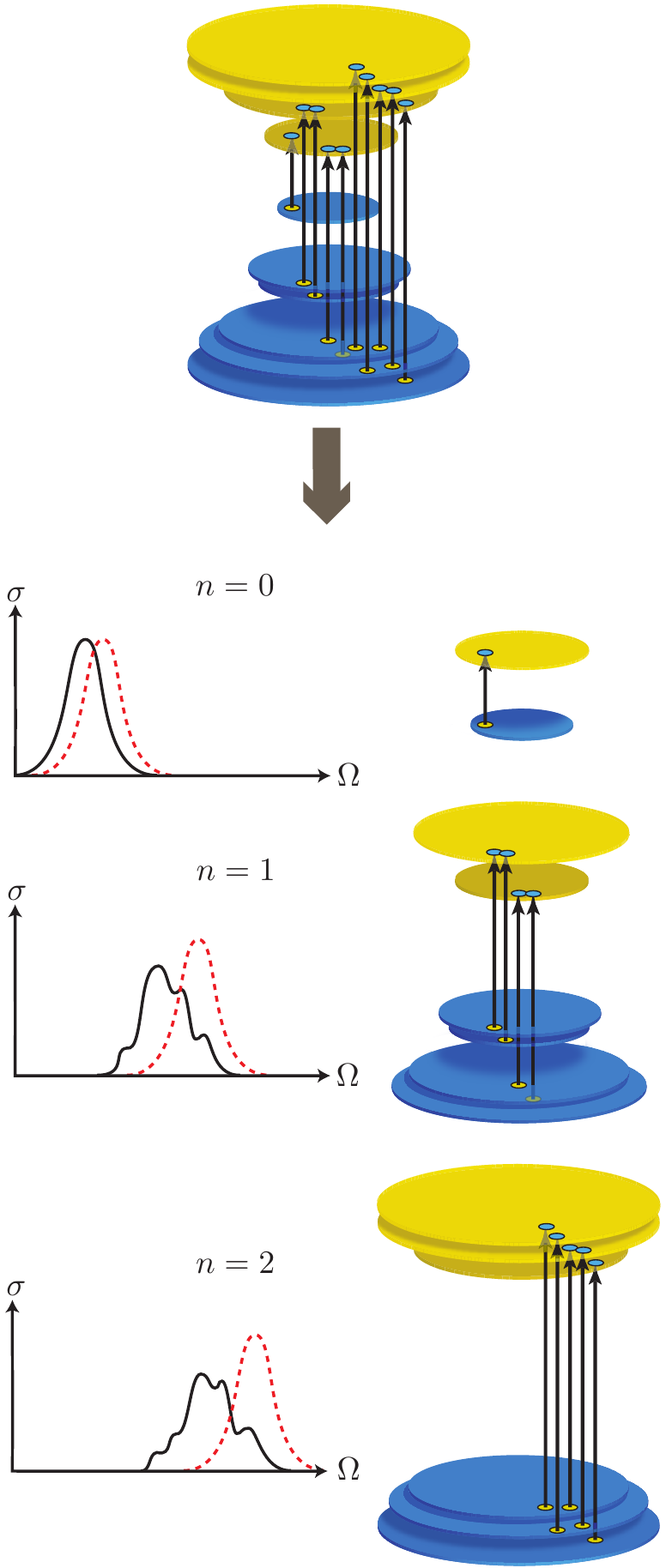}
\end{center}
\caption{(Color online) Schematic decomposition of the levels and transitions in the final diagram of Fig.~\ref{cones}. Sets of transitions are labeled by a non-negative integer $n$, corresponding to $\sigma^{(n)}$. For $n=0$, the set is made up of the transition $0\to1$. In the bare case, the members of each set for $n>0$ would be the transitions $-(n+1)\to n$ and $-n\to n+1$. In the dressed case, each of the additional levels or phonon-assisted peaks derives from a bare level, and the transitions involving it obey the same selection rules as those involving that associated level. To the left of each set, we show the absorption peaks arising from the transitions. In the bare case, the two involved transitions have the same energy, leading to a single peak; in the dressed case, all the involved transitions generically differ in energy and intensity, leading to a deformed set of peaks. \label{division}}
\end{figure}

In the dressed case the conductivity arises from the same four combinations of spectral functions, but each of those spectral functions is made up of several peaks rather than just one. This leads to more peaks in the conductivity, since there are a greater number of peaks to transition between. Appendix~\ref{selection_rules} describes the allowed transitions in detail. But for the moment, simply to orient the reader, we illustrate their general form in Fig.~\ref{division}, where the dressed levels and transitions shown in Fig.~\ref{cones}d are decomposed into schematic diagrams representing contributions to $\sigma^{(0)}$, $\sigma^{(1)}$, and $\sigma^{(2)}$. Next to each diagram, we show the corresponding bare (dashed red curves) and dressed (solid black) conductivity. For $n=0$, there is only one transition, $0\to1$, leading to a single absorption peak, which is shifted somewhat from the peak in the bare case. For $n=1$ and 2, additional peaks appear in the relevant spectral functions, leading to additional transitions and additional absorption peaks. Since the additional features in each $A_n$ are clustered together, the additional transitions, and therefore the additional absorption peaks, are likewise clustered. We will return to this diagram at the end of the following section, after having fully discussed the form of the dressed spectral functions. Readers interested in the final results for the optics but not the precise details that lead to them may skip to the final paragraph of Sec.~\ref{spectral_functions}, which summarizes the results and notation for the dressed spectral functions. 

\section{Effects of renormalization on the spectral functions}\label{spectral_functions}

\begin{figure}[tb]
\begin{center}
\includegraphics[width=\columnwidth]{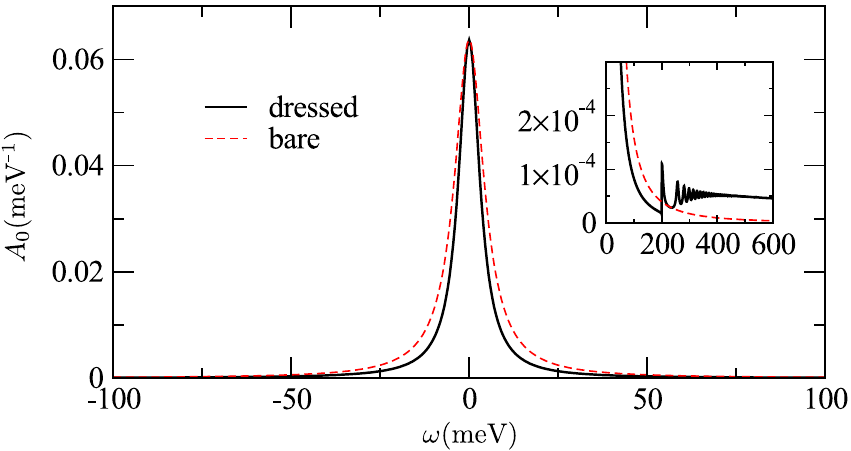}
\end{center}
\caption{(Color online) The spectral function $A_0(\omega)$ in the bare and dressed cases. While the peaks in the two cases have the same height, the dressed level is noticeably narrower, leading to lost spectral weight. The inset shows that this lost weight has gone into phonon-assisted peaks starting at the phonon frequency $\omega_E=200$meV (and the same below $-\omega_E$). Here $B=2.4$T, $\mu=0$, $A=500$meV, and $\Gamma=5$meV. \label{A0}}
\end{figure}

All the effects of electron-phonon coupling on each of the spectral functions 
$A_n$ can be understood from a transfer of spectral weight away from a single 
quasiparticle level into phonon-assisted peaks. This transfer is shown in 
Fig.~\ref{A0} for $A_0$ with a fairly weak magnetic field, $B=2.4$T, a fairly strong coupling, $A=500$meV, and $\mu=0$. We see 
that the dressed zeroth LL (shown in the solid black curve) has narrowed relative 
to the bare one (the dashed red curve), corresponding to a loss of weight in the 
level. The inset shows where this weight has been transferred: into an incoherent 
phonon-assisted region that begins at $|\omega|=\omega_E$.
The series of oscillations in this region are (approximately) at the energies 
$P_n$ where the peaks in $\Sigma$ occur, which, as discussed above, encode the 
LLs at energies shifted by $\omega_E$. In the limit of zero magnetic field, this 
region would become a smooth envelope of the oscillatory curve shown in
the inset, with an appearance akin to the incoherent phonon-assisted background 
of a conventional metal.

This transfer can easily be understood analytically. 
As discussed above (and in detail in 
Ref.~\cite{Carbotte:10}), for $|\omega|<\omega_E$, $\Sigma_1\simeq\Sigma_1(0)-
\lambda\omega$ and $\Sigma_2(\omega)=0$. If we restrict our attention to a quasiparticle 
peak that falls in this energy range and note that $\mu=\mu_0+\Sigma_1(0)$, it 
then follows immediately from Eq.~\eqref{dosA} that
the dressed spectral function describing that peak is given by
\begin{align}\label{weight_loss}
A_n(\omega)=\frac{1}{1+\lam}\frac{1}{\pi}\frac{\Gamma/(1+\lam)}{\left(\omega-\frac{M_n-\mu_0}{1+\lam}\right)^2+\left(\frac{\Gamma}{1+\lam}\right)^2}.
\end{align}
This is a simple Lorentzian, just as in the bare case, but shifted in position 
from $E_n=M_n-\mu_0$ to $E_n=(M_n-\mu_0)/(1+\lam)$, corresponding to an effective mass renormalization or renormalization of $v_F$. (We denote the energy of the $n$th Landau level relative to the Fermi energy by $E_n$ in both bare and dressed cases.) But while the bare peak has
a weight of 1, this dressed peak has its weight reduced
by a factor of $1/(1+\lam)$. The bare scattering rate $\Gamma$ is renormalized 
by the same factor. However, the height of the peak is unchanged:
at the position of the peak, $A_n\left(\omega=\frac{M_n-\mu_0}{1+\lam}\right)=\frac{1}{\pi\Gamma}$,
unrenormalized by the electron-phonon interaction---meaning the weight loss comes 
entirely from the renormalization of $\Gamma$. All of these approximate analytical 
results are confirmed by the full numerical results of Fig.~\ref{A0}.

\begin{figure}[tb]
\begin{center}
\includegraphics[width=\columnwidth]{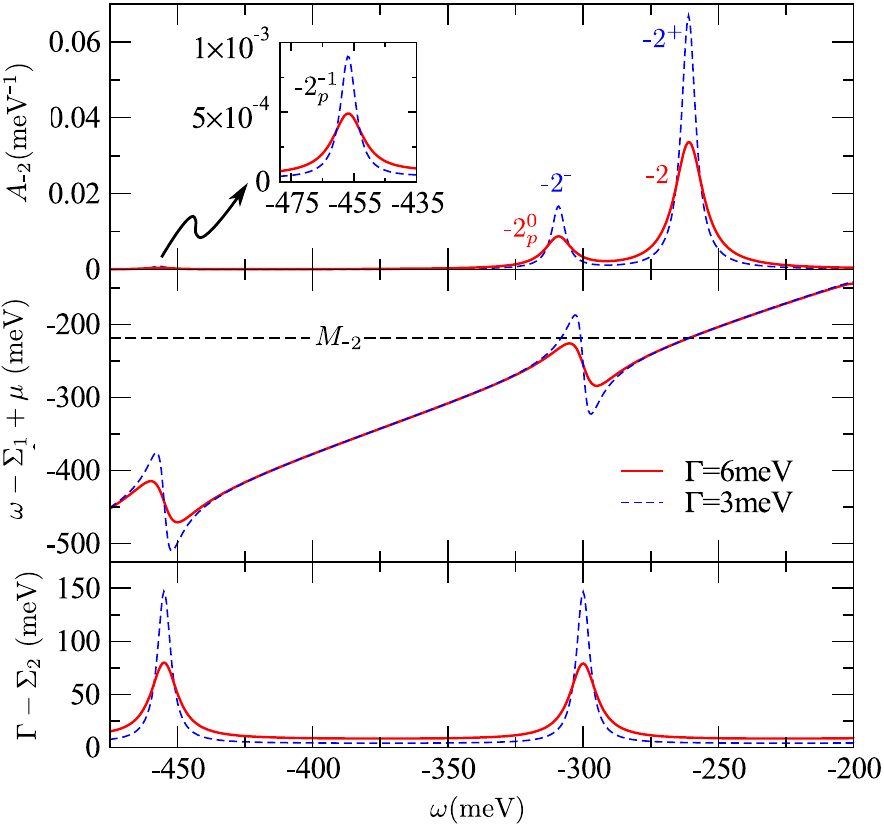}
\end{center}
\caption{(Color online) The spectral function $A_{-2}(\omega)$ (top frame) and the functions $\omega-\Sigma_1(\omega)+\mu$ (middle) and $\Gamma-\Sigma_2(\omega)$ (bottom) that determine it, each shown for two values of broadening $\Gamma$. In both cases, $B=18.2$T, $\mu_0=100$meV, $\omega_E=200$meV, and $A=250$meV. Also shown in the lower frame is $M_{-2}$ (horizontal dashed line). Peaks in the upper frame are either phonon-assisted (corresponding to peaks in $\Gamma-\Sigma_2$) or quasiparticle levels (corresponding to intersections of $\omega-\Sigma_1+\mu$ with $M_{-2}$); the number of intersections that occur is seen to depend on $\Gamma$. 
Landau levels are labeled with $n^\alpha$, where $\alpha=\pm$, while phonon-assisted peaks are labeled with $n_p^m$, where $m$ is the index on the energy $P_m$ of the corresponding peak in $\Gamma-\Sigma_2(\omega)$. The inset in the top frame shows that the peaks in $A_{-2}$ corresponding to those in $\Gamma-\Sigma_2$ are very weak when far from the largest peak in $A_{n}$. \label{A2}}
\end{figure}

Each $A_n$, for any integer $n$, shares this structure of a set of phonon-assisted peaks in addition to a main, somewhat depleted, coherent quasiparticle peak. These various types of peaks are illustrated in Fig.~\ref{A2}, which shows $A_{-2}$ in the upper frame and the functions that determine its behavior, $\omega-\Sigma_1(\omega)+\mu$ and $\Gamma-\Sigma_2(\omega)$, in the lower two frames.   Here we use a larger field, $B=18.2$T, to better separate the phonon-assisted peaks. We also use a weaker coupling, $A=250$meV; with $\omega_E=200$meV, this yields $\lambda\simeq 0.2$, which is realistic for graphene.\cite{Park:07}  Results for two scattering rates are shown, but for the moment, the reader should focus on the curves for the larger of the two, $\Gamma=6$meV, shown in solid red. Phonon-assisted peaks occur (in a rough sense) because of local minima in the function $\left|\omega-\Sigma_1(\omega)+\mu-M_n\right|$, which for $M_{n=-2}$ (shown as the horizontal dashed line) appear near $-310$meV and $-460$meV in the solid curve in the middle frame. Evidently the positions of these peaks will depend on $n$, but they always occur near the energies $P_m=M_m-\mu_0\pm\omega_E$, where oscillations in $\Sigma_1(\omega)$ and peaks of $\Sigma_2(\omega)$ occur. Hence, we label their energies as $P_{m,n}$, and we label the phonon-assisted peaks themselves as $n_p^m$. The two phonon-assisted peaks appearing in the solid red curve in the top frame are labeled $-2_p^0$ and $-2_p^{-1}$, indicating that they are associated with $A_{-2}$ and with the peaks in the self-energy at $P_0$ and $P_{-1}$. The latter of these, shown in an inset that magnifies the region around $-450$meV, is reduced by an order of 100 relative to the larger one. The larger one is of significant weight, and in fact, the weight of the quasiparticle peak in this case is reduced by more than a factor of $1/(1+\lam)$; as we will show in the next section, quasiparticle peaks are generally reduced below the $1/(1+\lam)$ weight when they lie near a peak in the self-energy.

Because of the oscillations in the self-energy introduced by the transfer of spectral weight, the quasiparticle level itself may be modified beyond narrowing: it may be split. We will define any peak to be a quasiparticle level if it lies at an energy $\omega=E_n^\alpha$ that gives a zero for the real part in the denominator of the defining equation for $A_n(\omega)$, Eq.~\eqref{dosA}; that is, it must satisfy
\begin{equation}
E^\alpha_n-\Sigma_1(E^\alpha_n)+\mu=M_n,\label{E_n}
\end{equation}
where the index $\alpha$ accounts for multiple solutions to Eq.~\eqref{E_n}. We label these peaks as $n^\alpha$. In the bare case, Eq.~\eqref{E_n} reduces to $E_n=M_n-\mu_0$, and there is a single peak. For the solid red curves in Fig.~\ref{A2}, there is likewise only one such peak, labeled with a $-2$, corresponding to the lone intersection of the solid curve in the middle frame with $M_{-2}$; as we would expect, this peak occurs very near $(M_{-2}-\mu_0)/(1+\lam)$. But the curve for $\Gamma=3$meV (shown in dashed blue) in the middle frame intersects $M_{-2}$ a second time, at the upper left side of the oscillation, meaning that the peak in the dashed curve in the upper frame at this energy is classified as a quasiparticle level, rather than as a phonon-assisted peak as it was for $\Gamma=6$meV. In this case the $n=-2$ level is split into the two substituent peaks labeled $-2^\pm$. (Note that the third intersection, at around $-300$meV on the right-hand side of the crest of the oscillation, does not correspond to a peak in $A_{-2}$, because $\Gamma-\Sigma_2$ spikes in this region, as shown in the bottom frame.) The weight of the peak $2^-$ is identical to that of the phonon-assisted peak $2_p^0$ in the $\Gamma=3$meV case.

From this, we see that the classification of peaks depends very finely on the choice of parameters---even if the parameter in question does not shift the distribution of weight from one peak to another. Equation~\eqref{E_n} is chosen to define a peak as a quasiparticle level because it defines the usual energy of a dressed quasiparticle damped in its motion by the imaginary part  of the self-energy, $\Gamma-\Sigma_2(E^\alpha_n)$. But the distinction between a split level versus a level plus a phonon-assisted peak is obviously physically fuzzy, though mathematically sharp. However, the spectral weight of a phonon-assisted peak is generally less than $\lam/(1+\lam)$ (the minimum weight lost by the associated quasiparticle), which is small for graphene. Furthermore, regardless of a peak's classification, we can say that in general, the further a peak lies from $(M_n-\mu_0)/(1+\lam)$, the smaller its weight, as is seen in the strongly reduced weight of the phonon-assisted peak in the inset of Fig.~\ref{A2}. So any significant phonon-assisted peak will lie near there and will not radically shift the spectral weight away from the expected energy of the dressed quasiparticle.


\begin{figure}[tb]
\begin{center}
\includegraphics[width=\columnwidth]{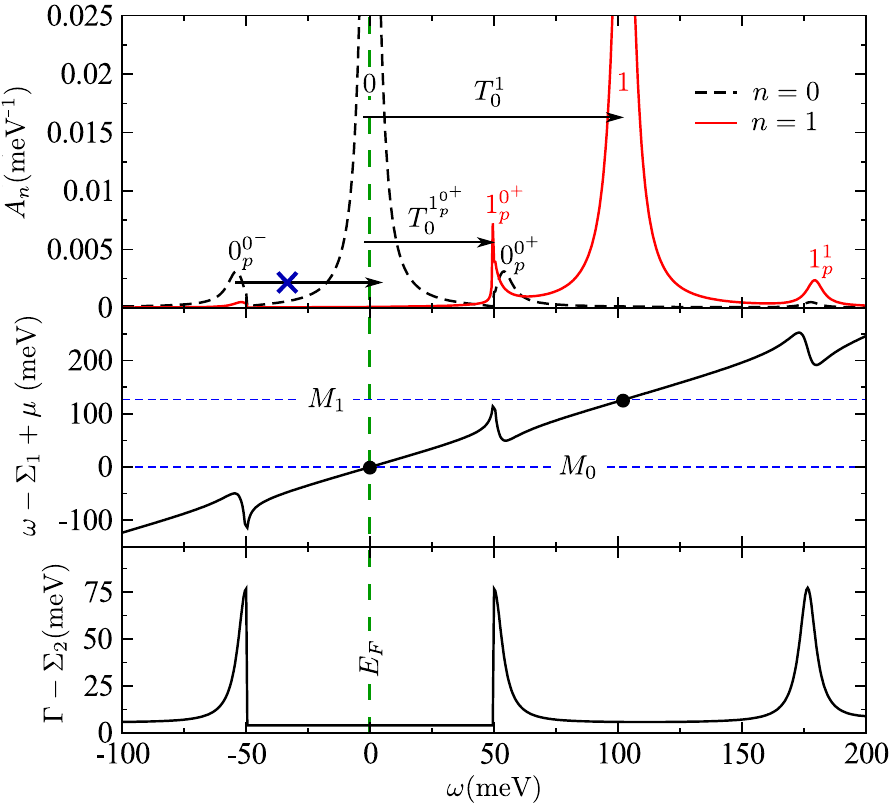}
\end{center}
\caption{(Color online) Two spectral function $A_0(\omega)$ and $A_1(\omega)$ (top frame) and the transitions between them, together with the functions $\omega-\Sigma_1(\omega)+\mu$ and $\Gamma-\Sigma_2(\omega)$ (lower two frames), which determine their form. The Fermi energy, which must lie between the initial and final state, is indicated by the dashed green vertical line. Examples of allowed transitions are indicated by arrows; an example of a disallowed transition is crossed out with an {\sf X}. Solid black circles mark intersections of $\omega-\Sigma_1+\mu$ with $M_0$ and $M_1$, where quasiparticle peaks occur in $A_0$ and $A_1$, respectively. Here $\omega_E=50$meV, $B=12.2$T, $\Gamma=4$meV, and $A=250$meV.\label{A0A1}}
\end{figure}

With the basic form of the spectral functions understood, we now turn to the manner in which they interact to yield the conductivity. As previously stated, one $A_n(\omega)$ describes the initial level and the other describes the final level involved in the absorption of a photon of energy $\Omega$. And as also stated previously, the absolute values of the two levels must differ by one: for intraband transitions, either $n\to n+1$ or $-(n+1)\to-n$; for interband, $-n\to n+1$ or $-(n+1)\to n$ (where $n$ is a non-negative integer in all cases). More details about the selection rules are presented in Appendix~\ref{selection_rules}. In Fig.~\ref{A0A1}, we illustrate the form of the transitions from $A_0$ to $A_1$, with the top frame showing the two spectral functions, and the lower two frames showing the functions ($\omega-\Sigma_1+\mu$ and $\Gamma-\Sigma_2$) that determine them. Optical selection rules allow transitions from any peak in $A_0(\omega)$ (dashed black curve) that falls below the Fermi energy (vertical dashed line) to any peak in $A_1(\omega)$ (solid red curve) that falls above the Fermi energy. The two most prominent of these transitions (i.e., the two with the largest optical weight) are indicated by the black horizontal arrows, with the labels $T_{n_i}^{n_f}$ on the arrows denoting a transition from an initial peak $n_i$ to a final peak $n_f$. Black-curve-to-black-curve or red-to-red transitions are forbidden, as indicated in the figure by a crossed-out arrow. Here we have used an unrealistically small value of the phonon frequency, $\omega_E=50$meV, in order to display several phonon-assisted peaks for each spectral function. In the lower two frames, we see oscillations in $\Sigma_1$ and peaks in $\Sigma_2$ at $\omega=P_{0^\pm}=M_0\pm\omega_E=\pm50$meV, and at $P_1=M_1+\omega_E\simeq175$meV (here $\mu_0=0$). Note that the peak $1_p^{0^+}$ is larger than $0_p^{0^+}$ because the former lies closer to $M_1/(1+\lam)$ than the latter does to $M_0/(1+\lam)$. Also note that these two peaks lie at slightly different positions; that is, $P_{0^+,0}\neq P_{0^+,1}$, though both are within $\Gamma$ of $P_{0^+}$.

We are now positioned to fully understand Fig.~\ref{division}. The levels arranged in cones in that picture roughly depict the full spectral function $A(\vec k, \omega)$, which contains a sum over the individual $A_n(\omega)$'s. In that case, for a given $m$ the phonon-assisted peaks at energies $P_{m,n}$ all add together to create a single peak at $P_m$ (which we referred to as a ``phonon peak'' in the DOS\cite{Pound:11a,Pound:11b}). For simplicity, assume that only one phonon-assisted peak near a given $P_m$, if any, contributes significant weight to that sum. Now, for $\sigma^{(0)}$ the transition is from the zeroth
Landau level to the first, leading to a single absorption peak, shifted down in energy from the bare peak due to the rescaling of the LLs by $1/(1+\lam)$. 
For $\sigma^{(1)}$, the transitions are from peaks in $A_{-1}$ to peaks in $A_2$ or from peaks in $A_{-2}$ to peaks in $A_1$. Each of the inital states, corresponding to $A_{-1}$ or $A_{-2}$, are either split into two or have a significant weight transferred into a single phonon-assisted peak, while the final states corresponding to $A_2$ or $A_1$ each have only one significant peak.
This leads to four distinct transitions and hence four distinct, though closely clustered, structures in the
conductivity, as shown schematically on the lefthand side of the figure. Finally, for
$\sigma^{(2)}$, the transitions are from peaks in $A_{-2}$ to peaks in $A_3$ or from peaks in $A_{-3}$ to peaks in $A_2$.
Here one of the initial states---$A_{-2}$---and one of the final states---$A_3$---each have two significant peaks, while the other initial and final states each have one. This provides five distinct transitions and five absorption peaks. Note that because all the features in the spectral functions are discretely spaced, the equivalent of a Holstein process can be understood as a transition between a quasiparticle level and a phonon-assisted peak that appears as a line in the spectral function, rather than requiring the image of an emitted phonon as in Fig.~\ref{cones}b.

In summary, each spectral function $A_n(\omega)$ has two types of peaks: quasiparticle peaks labeled $n^\alpha$ at energies $E_n^\alpha$ satisfying $E_n^\alpha-\Sigma_1(E_n^\alpha)+\mu-M_n=0$; and phonon-assisted peaks labeled $n_p^m$ at energies $P_{m,n}\simeq P_m=M_m-\mu_0\pm\omega_E$. That is, we can write the approximate expression
\begin{equation}
A_n(\omega) = \sum_\alpha W_n^\alpha\delta(\omega-E^\alpha_n)+\sum_{m=-\infty}^\infty W_{m,n}\delta(\omega-P_{m,n}),\label{A_n model}
\end{equation}
where the spectral weights $W_n^\alpha$ and $W_{m,n}$ may range from 0 to 1. The weight of a peak of either type is generally large if the peak lies near $(M_n-\mu_0)/(1+\lam)$, the energy of a dressed quasiparticle given a simple effective mass renormalization, and small if it lies far from that energy. A transition from an initial peak $n_i$ to a final peak $n_f$ is denoted $T_{n_i}^{n_f}$. This notation will be used throughout the remainder of the paper.


\section{General behavior of absorption peaks}\label{general behavior}

\begin{figure}[tb]
\begin{center}
\includegraphics[width=\columnwidth]{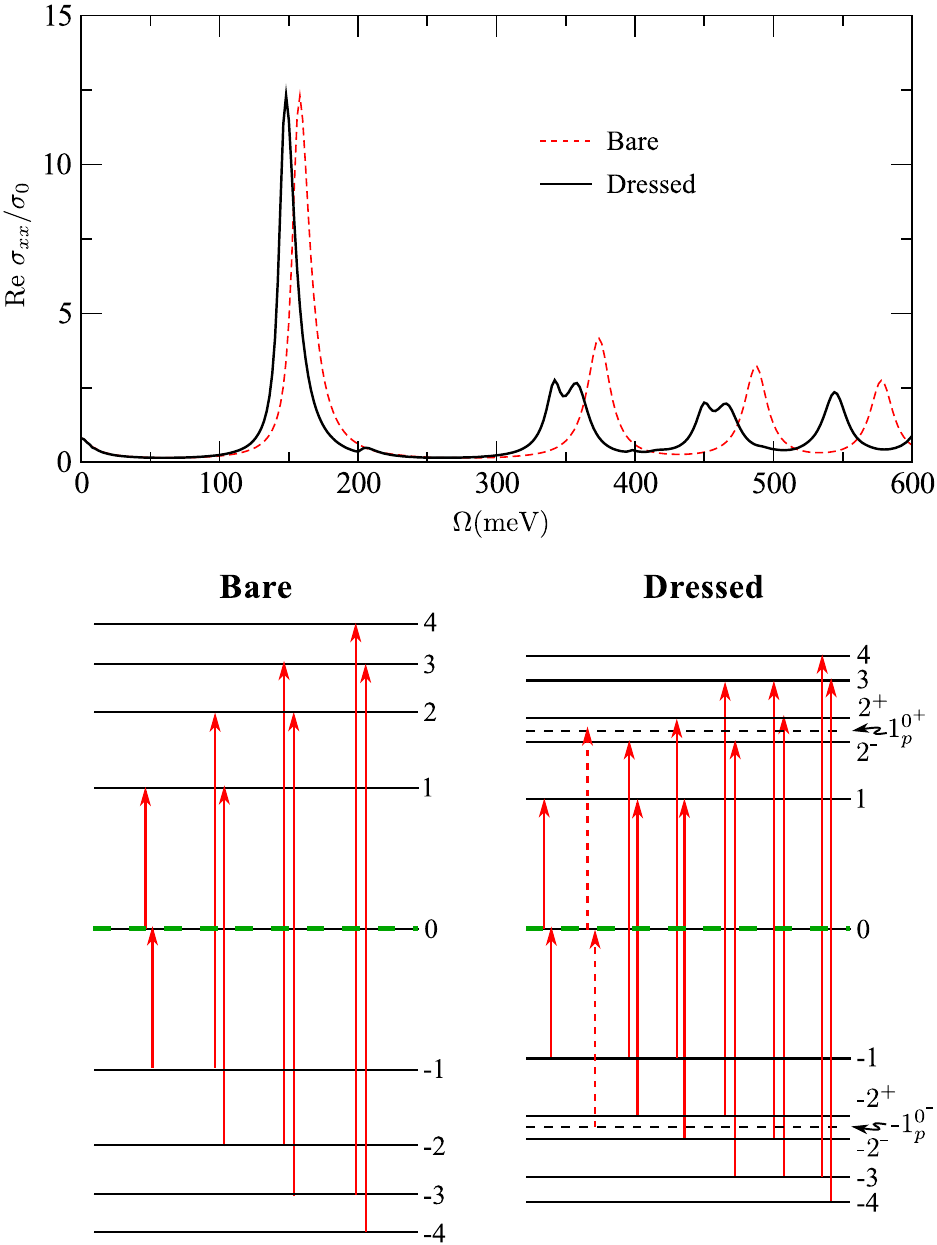}
\end{center}
\caption{(Color online) Upper frame: the real part of the diagonal conductivity in the bare (dashed red curve) and dressed (solid black) cases for a small coupling $A=80$meV. The other parameters are $B=18.2$T, $\mu=0$, $\Gamma=5$meV, and $\omega_E=200$meV. Lower frame: the level diagrams for the two cases with all significant transitions between them. The thick, dashed green line shows the position of the Fermi energy. The dotted lines indicate phonon-assisted peaks rather than quasiparticle levels.\label{lowA}}
\end{figure}

With the form of the spectral functions understood, we may now understand the form of the conductivity. In the top frame of Fig.~\ref{lowA}, we show results for the real part of the diagonal conductivity, $\sxx$, for $\mu=0$ and the fairly weak coupling $A=80$meV, which allows us to clearly see renormalization effects. We see four distinct structures in both the bare (dashed red curve) and dressed (solid black) cases, corresponding to $\sxx^{(0)}$, $\sxx^{(1)}$, $\sxx^{(2)}$, and $\sxx^{(3)}$. As expected, each structure is a simple Lorentzian in the bare case. Compared to these bare lines, the dressed structures are shifted down in energy, with the second and third lines being split in two. A small bump also appears near $\Omega=\omega_E=200$meV. A detailed energy level diagram describing the transitions giving rise to these various
peaks is shown in the lower frame of the figure. On the left is the bare
level scheme; on the right, the dressed one. The heavy dashed horizontal green line is the Fermi energy. Because the Fermi energy lies in the middle of the zeroth LL at $\mu=0$, all transitions come in pairs: $-1\to 0$ and $0\to 1$, $-1\to 2$ and $-2\to 1$, etc. The split absorption peaks in $\sxx^{(1)}$ and $\sxx^{(2)}$ in the dressed case come from the splitting of the $\pm 2$ LLs into $+2^{\pm}$ and $-2^\pm$. Between each of the two split levels, near the energies $P_{0^\pm}$, lie phonon-assisted peaks $1^{0^+}_p$ and $-1^{0^-}$, shown as horizontal
dashed (black) lines; the two transitions involving these peaks lead to the small Holstein-like structure seen
at $\Omega\simeq 200$meV.

Beyond the basic understanding afforded by the level diagrams, we can also analytically derive some of the features of the dressed lines. In particular, the absorption peaks are not only shifted down in energy, but as is apparent in the first undivided peak (i.e., $\sxx^{(0)}$, around 150meV), they are also narrowed relative to the bare peaks, while their height is unchanged. The first peak arises from transitions between LLs 0 and $\pm1$. For the parameters used here (given in the caption), these LLs lie far from any peaks in the self-energy. In such cases, we can straightforwardly continue the reasoning used to derive Eq.~\eqref{weight_loss}. Consider an intraband transition $T_n^{n+1}$ where the quasiparticle peaks $n$ and $n+1$ are well separated from any peak in the self-energy. From Eq.~\eqref{sigma_xx}, this transition's contribution to $\sxx^{(n)}(\Omega)$ is given by
\begin{equation}
\sxx^{(n)}(\Omega) = \frac{\sigma_0 M_1^2}{\Omega}\int_0^\Omega d\omega A_{n+1}(\omega)A_n(\omega-\Omega).
\end{equation}
(Note that this may not be the whole of $\sxx^{(n)}(\Omega)$, which can also contain contributions from interband transitions $T_{-n}^{n+1}$ and $T_{-(n+1)}^{n}$.) As in the discussion surrounding Eq.~\eqref{weight_loss}, this can be simply evaluated in the dressed case by replacing each $A_n(\omega)$ with $A_n\left[\omega(1+\lambda)\right]$, which in the $\Gamma\to0$ limit leads to
\begin{equation}
\sxx^{(n)}(\Omega) = \frac{\sigma_0 M_1^2}{\Omega(1+\lambda)}\delta\left[M_{n+1}\!-M_n\!-\Omega(1+\lambda)\right].
\end{equation}
This returns the first line of Eq.~\eqref{sigmaxx_bare} in the bare case, where $\lam=0$. Evaluating $\Omega(1+\lambda)$ at the value determined by the $\delta$-function, restoring a small, constant half-width $2\Gamma$ to the $\delta$-function (corresponding to the half-width $\Gamma$ in each $A_n$), and making a trivial rearrangement, we arrive at
\begin{align}
\sxx^{(n)}(\Omega) &= \frac{\sigma_0M_1^2}{M_{n+1}-M_n}\nonumber\\
&\quad\times\frac{1}{1+\lambda}\frac{1}{\pi}\frac{2\Gamma/(1+\lambda)}{\left(\Omega-\frac{M_{n+1}-M_n}{1+\lam}\right)^2+\left(\frac{2\Gamma}{1+\lambda}\right)^2}.\label{dressed_peak}
\end{align}
Here we see that relative to the bare case, the position of the peak has been shifted from $M_{n+1}-M_n$ to $(M_{n+1}-M_n)/(1+\lam)$, the width has been decreased from $2\Gamma$ to $2\Gamma/(1+\lam)$, and the weight has been decreased by the overall factor of $1/(1+\lam)$. The height of the peak is, however, unchanged by renormalization: 
\begin{equation}
\sxx^{(n)}\left(\Omega=\frac{M_{n+1}-M_n}{1+\lam}\right)=\frac{\sigma_0}{2\pi\Gamma}\frac{M_1^2}{M_{n+1}-M_n}.\label{dressed_height}
\end{equation}
Both this constancy of height and the rescalings of peak position and width by $1/(1+\lam)$ are verified by the full numerical results for the first peak in Fig.~\ref{lowA}. And they would also hold true if we performed the calculation for interband transitions between two LLs that are well separated from any peaks in the self-energy.

\begin{figure}[tb]
\begin{center}
\includegraphics[width=\columnwidth]{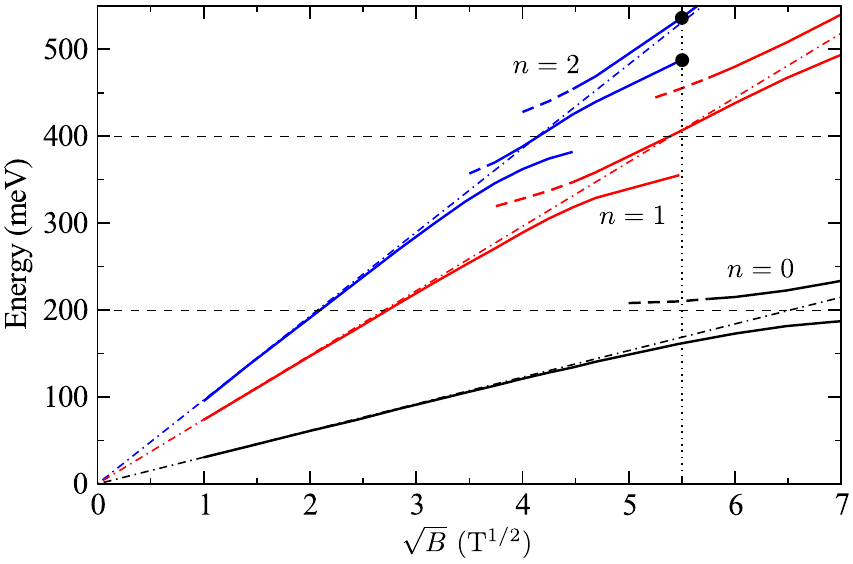}
\end{center}
\caption{(Color online) Transition energies associated with peaks in $\sxx^{(0)}$, $\sxx^{(1)}$, and $\sxx^{(2)}$, as a function of $\sqrt B$, for parameters $\mu=0$, $\Gamma=5$meV, $\omega_E=200$meV, $A=250$meV. Only those transitions that give rise to significant absorption peaks are shown. Solid curves correspond to transitions between levels; dashed curves, to transitions between levels and phonon-assisted peaks; and the dot-dashed lines, to $(M_{n_f}-M_{n_i})/(1+\lam)$, where $n_i$ is the initial level and $n_f$ is the final level. The dotted horizontal lines mark $\omega_E$ and $2\omega_E$.\label{transition_energies}}
\end{figure}

The energies of the levels shown in the bottom frame of Fig.~\ref{lowA}
depend, of course, on the magnitude of the applied magnetic field $B$.
For the bare case, these are $M_n$ and $M_{-n}$, which are both
proportional to the square root of $B$ (i.e., $\sqrt{B}$). The $B$-dependence 
in the dressed case is illustrated in Fig.~\ref{transition_energies}
for the first three sets of transition energies, namely, $n=0$, 1, and 2, corresponding to peaks in $\sxx^{(0)}$, $\sxx^{(1)}$, $\sxx^{(2)}$.
Here $\mu$ remains zero, but $A=250$meV in order to accentuate the effects of coupling.
The solid curves correspond to transitions between quasiparticle levels, and
the dashed curves, to transitions between a quasiparticle level and a phonon
peak. The dot-dashed lines give $(M_{n_f}-M_{n_i})/(1+\lambda)$, where
$n_f$ and $n_i$ are the final and initial states, respectively; this corresponds to the simple effective mass renormalization.
The dotted horizontal lines mark $\omega_E=200$meV and $2\omega_E=400$meV. Away from these
energies, the energies of the optical absorption peaks follow closely the
$\sqrt{B}$ law and lie close to the dot-dashed curves. So in this
regime, the simple effective mass renormalization applies. In the vicinity 
of the energies $\omega_E$ or $2\omega_E$, the situation is more
complex, as these energies correspond to the main quasiparticle peak in one or both of $A_{n_i}$ or $A_{n_f}$ lying near $\pm\omega_E$, leading to a disruption of their behavior, with a significant amount of their weight being transferred into phonon-assisted peaks
or being split between two quasiparticle levels.

\begin{figure}[tb]
\begin{center}
\includegraphics{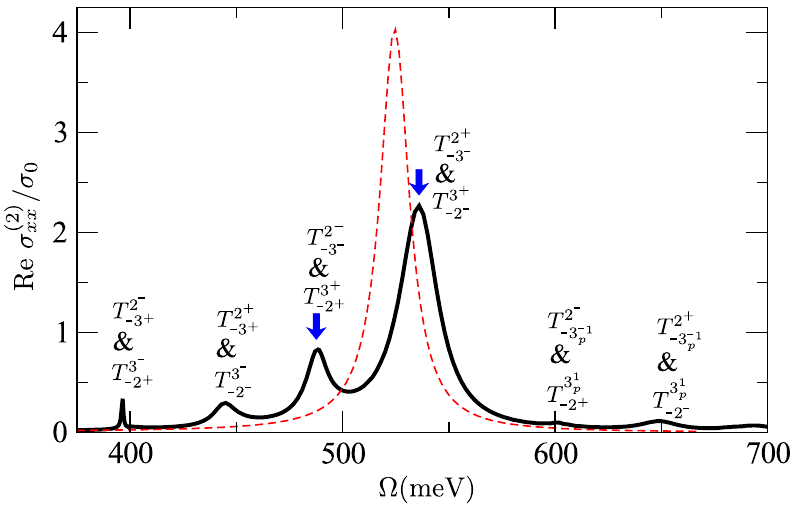}
\end{center}
\caption{(Color online) The $n=2$ piece of the diagonal conductivity (solid black curve) as a function of frequency at $B=5.5^2$T, with parameters as in Fig.~\ref{transition_energies}. Each peak is labeled with the transitions giving rise to it. The dashed red curve is the bare conductivity shifted down by a factor of $1/(1+\lam)$. Blue arrows point to the two absorption peaks corresponding to the solid circles on the $n=2$ curves in Fig.~\ref{transition_energies}.\label{sigma_n=2}}
\end{figure}

In Fig.~\ref{transition_energies} we have shown only the significant transitions, ignoring any transition leading to an absorption peak with weight less than 5\% of the maximum possible for the particular $\sxx^{(n)}$. The kinds of peaks that have been ignored are illustrated in Fig.~\ref{sigma_n=2}, which shows $\sxx^{(2)}(\Omega)$ (solid black curve) for a single value of magnetic field, $B=5.5^2$T; this value of $B$ is marked by a vertical dotted line in Figure~\ref{transition_energies}. We note six peaks in Fig.~\ref{sigma_n=2}, four of which arise from transitions between two split levels (as labeled in the figure), and two of which arise from transitions involving phonon-assisted peaks. The two most prominent, indicated with arrows,
correspond to the transition energies marked by solid circles in Fig.~\ref{transition_energies}, while the others
are too weak to appear in Fig.~\ref{transition_energies}.
The dashed red curve shows the bare case shifted down by a factor of $1/(1+\lam)$, and we see that the optical weight in $\sigma^{(2)}_{xx}$ is clustered around the peak in this curve, with the most prominent peak in $\sigma^{(2)}_{xx}$ falling very near to it. From this, we can surmise that transition energies falling nearest to the dot-dashed straight lines in Fig.~\ref{transition_energies} are most prominent.

\begin{figure}[tb]
\begin{center}
\includegraphics{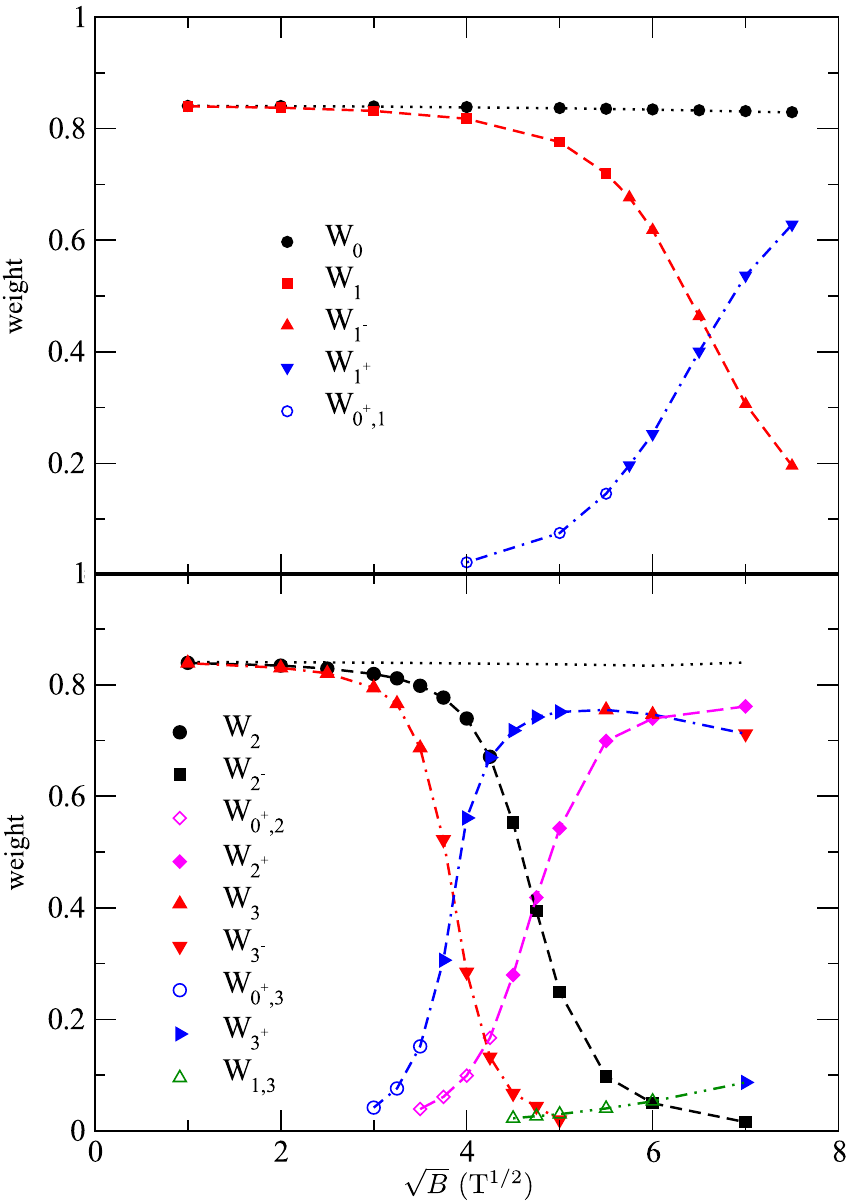}
\end{center}
\caption{(Color online) The weights of all significant peaks in $A_0(\omega)$, $A_1(\omega)$, $A_2(\omega)$, and $A_3(\omega)$, with parameters as in Fig.~\ref{transition_energies}. The upper frame shows weights of peaks in $\sxx^{(0)}$; the lower frame, those involved in $\sxx^{(2)}$. Open symbols correspond to phonon-assisted peaks, while solid ones correspond to quasiparticle levels. The dotted black line in both cases represents $1/(1+\lam)$. Other lines connecting symbols are to guide the eye.\label{weights}}
\end{figure}

That notion is confirmed by Fig.~\ref{weights}, which makes more precise the result of Sec.~\ref{spectral_functions} that the peaks in the spectral functions are more heavily weighted the closer they lie to the simple effective mass renormalization prediction. The figure shows the weights of all significant peaks in $A_0$, $A_1$, $A_2$, and $A_3$ as functions of $\sqrt{B}$ for the same parameters as used in Fig.~\ref{transition_energies}. The notation here follows that introduced by Eq.~\eqref{A_n model}. Since the optical weight in a given absorption peaks is simply the product of the weights in the initial and final peaks, these plots are sufficient to predict and understand the weights in the absorption peaks. (Because $\mu=0$ here, the levels are symmetric about $\omega=0$, even in the dressed case, so the weights of the peaks for $A_{n<0}$ are not necessary.) For example, the products obtainable from the upper frame determine the weights of peaks in $\sxx^{(0)}$: $W_0W_1$ gives the weight of the transition indicated by the lower solid black curve below about $\sqrt{B}=5.5T^{1/2}$ in Fig.~\ref{transition_energies}; $W_0W_{1^+}$, the lower solid black curve above that field value; $W_0W_{0^-,1}$, the dashed black curve; $W_0W_{1^-}$, the upper solid black curve. Because the $n=0$ LL is fixed at $\omega=0$ for this case of $\mu=0$, its weight $W_0$ (indicated by solid black circles) has the maximal value of $1/(1+\lam)$ (indicated by a dotted black curve in both frames). The weights of the transitions are hence governed by the changing weights in $A_1$. As $B$ increases, the principal peak in $A_1$ approaches $\omega_E$, causing its weight to transfer into a phonon-assisted peak near there.  As $B$ increases further, the phonon-assisted peak transitions into a quasiparticle level, meaning the principal peak has been split in two. At this point, around $\sqrt B = \sqrt{5.5}T^{1/2}$, the original peak has begun to deviate from the simple renormalization value $M_1/(1+\lam)$ but still lies very near it, as we can infer from the proximity of the lower solid black curve to the dash-dotted black line in Fig.~\ref{transition_energies}, and it contains most of the spectral weight. The more $B$ increases, the more weight is transfered from the original quasiparticle level into the additional one, as the old peak deviates more from $M_1/(1+\lam)$ and the additional peak gets closer to that value. Although the situation is more complicated for $\sxx^{(1)}$ and $\sxx^{(2)}$, since none of the principal peaks are pinned at one position and weight in those cases, we can still track the same behavior by correlating the weights in the lower frame of Fig.~\ref{weights} with the $n=2$ set of curves in Fig.~\ref{transition_energies}.


\begin{figure}[tb]
\begin{center}
\includegraphics{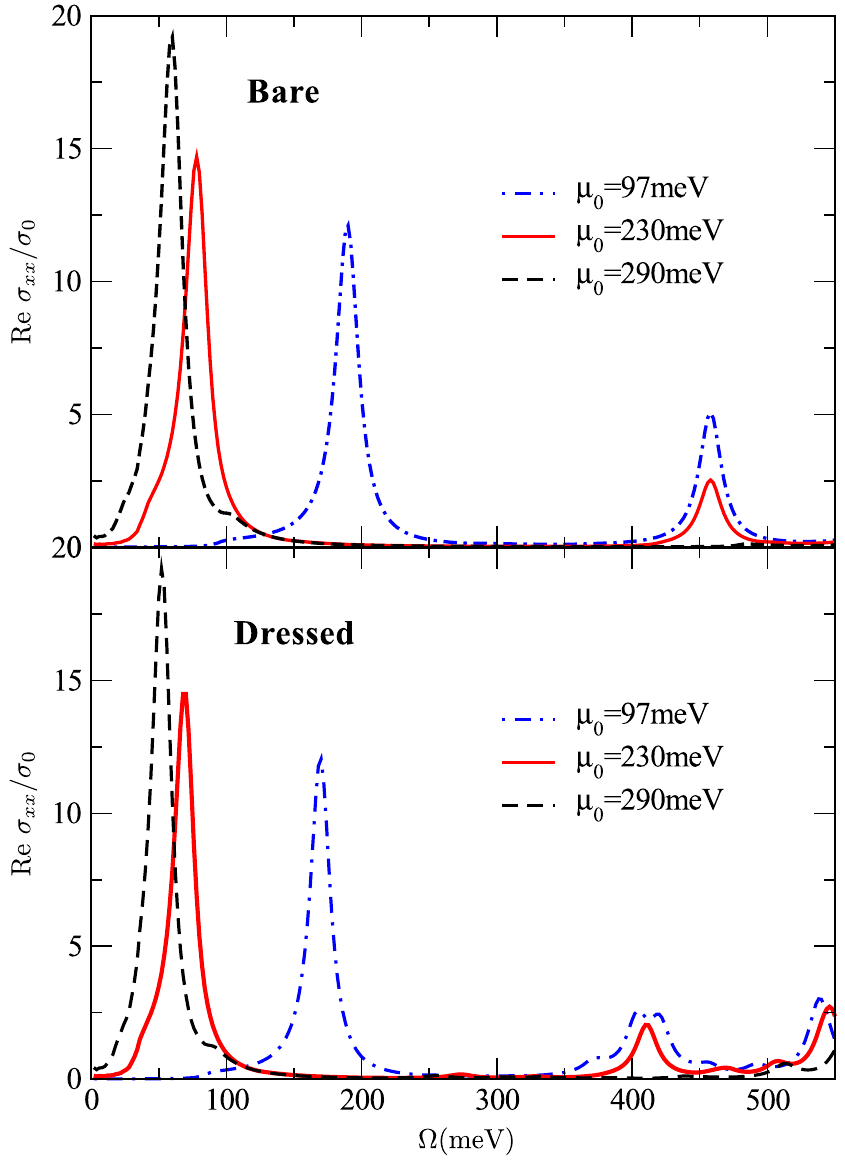}
\end{center}
\caption{(Color online) Effect of increasing chemical potential on the conductivity. The three values of $\mu_0$ are chosen such that the Fermi energy lies between the 0th and 1st LL, the 1st and 2nd, and the 2nd and 3rd, respectively. The remaining parameters are $B=27.4$T, $\Gamma=5$meV, $\omega_E=200$meV, and $A=150$meV.\label{varmu}}
\end{figure}

\section{Varying the chemical potential}\label{chemical potential}

\begin{figure}[tb]
\begin{center}
\includegraphics{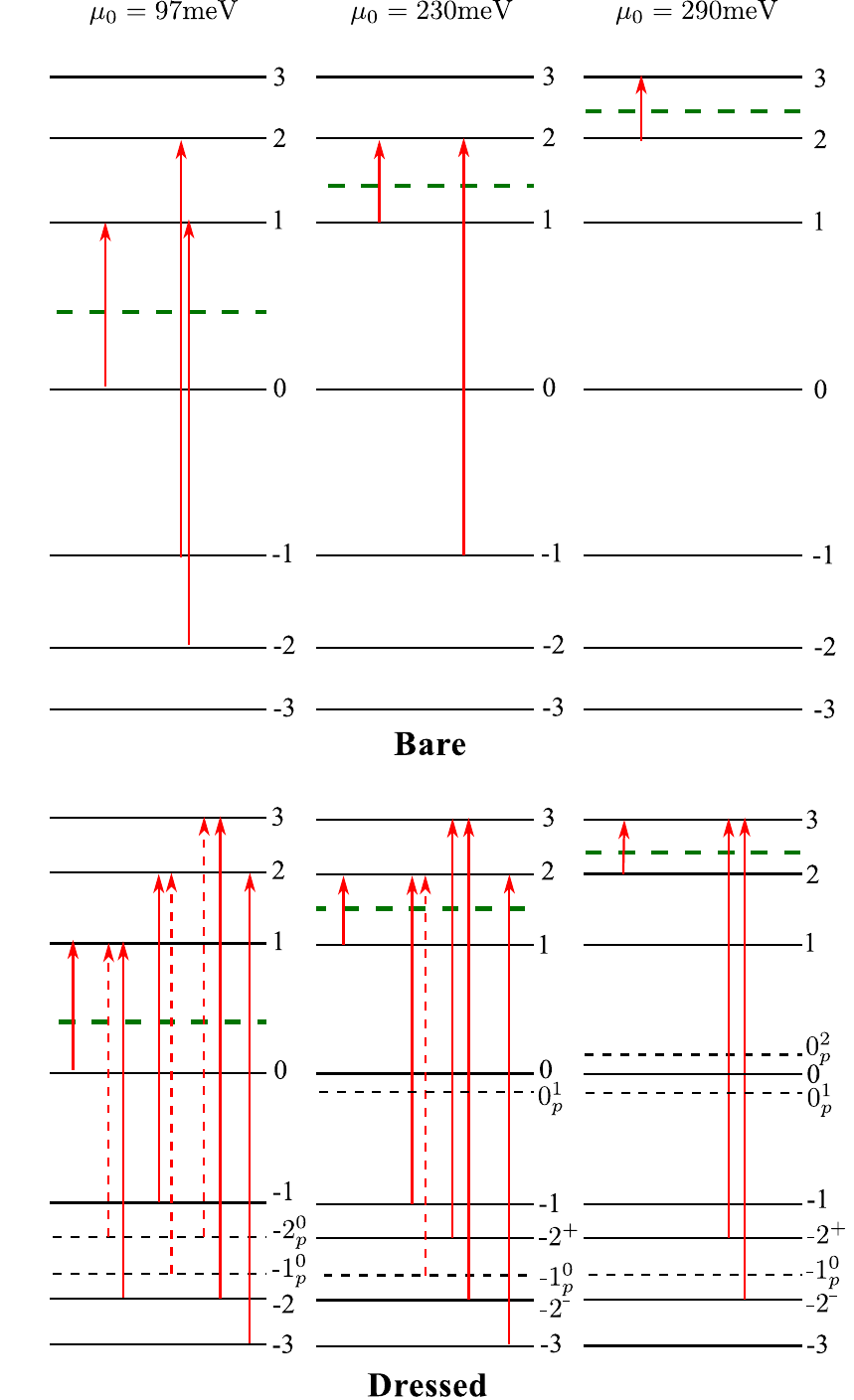}
\end{center}
\caption{(Color online) The level diagrams corresponding to Fig.~\ref{varmu}. As $\mu_0$ is increased, the Fermi energy (indicated by the dark, dashed, green line) moves up through the levels, causing the lowest-energy, intraband transition to decrease in energy as the level spacing at the Fermi energy decreases, and progressively eliminating interband transitions as more levels in the upper band become occupied.\label{varmu_levels}}
\end{figure}

Used as an element in a field effect device, graphene can be gated
to increase the charge imbalance, i.e., to vary the bare chemical
potential over a significant range. When this is done, the observed
pattern of optical lines is modified in a very specific way. Figure~\ref{varmu} shows this behavior in both the bare (upper frame) and dressed (lower frame) cases.
Three values of $\mu_0$ are used: 97meV (dash-double-dotted), 230meV (solid),
and 290meV (dashed), which are chosen to fall respectively between the zeroth and first Landau
levels, the first and second, and the second and third. In the bare case, which has been studied previously,\cite{Gusynin:07b} two types of peaks are visible. First, there is a low-frequency peak arising from an intraband transition. As $\mu_0$ is increased, this peak shifts to lower frequency. Second, there is a peak at $\simeq460$meV, which arises from interband transitions. Unlike the low-frequency peak, this peak does not shift position as $\mu_0$ varies. Instead, when $\mu_0$ is increased from 97meV to 230meV, the peak is halved in intensity, and when $\mu_0$ is further increased to 290meV, the peak disappears entirely. The behavior of both types of peaks can be understood from the level diagrams shown in the upper frame of Fig.~\ref{varmu_levels}. There, we see that for $\mu_0=97$meV, the intraband transition giving rise to the low-frequency absorption peak is $T^1_0$. As $\mu_0$ increases to 230meV and then to 290meV, the relevant intraband transition becomes $T_1^2$ and then $T_2^3$, which are at increasingly low frequencies. We also see that the visible interband absorption peak is due to the transitions $T_{-2}^1$ and $T_{-1}^2$, which have the same energy. As $\mu_0$ increases from 97meV to 230meV, the $n=1$ LL falls below the Fermi energy and the $T_{-2}^1$ transition becomes Pauli blocked; since one of the two transitions ceases, the absorption peak is halved in intensity. When $\mu_0$ is further increased to 290meV, the $n=2$ level falls below the Fermi energy, leading to the absorption peak disappearing entirely.

This general behavior---the intraband absorption peak decreasing in energy and the interband ones losing weight as the chemical potential is increased---remains largely unaltered by the electron-phonon interaction. In the lower frame of Fig.~\ref{varmu}, we see that in the dressed case, the absorption peaks from intraband transistions behave just as in the bare case, only with each peak shifted down by the factor $1/(1+\lam)$. This is to be expected, since the intraband transitions occur between levels near the Fermi energy and are hence unaffected by the phonon effects that begin for $|\omega|\geq\omega_E$. Because all peaks are roughly shifted down by $1/(1+\lam)$, an interband peak not visible in the bare case in this energy range now appears to the far right, but we again focus on the first interband ``peak", which now appears as a richer structure at $\simeq410$meV. For $\mu=97$meV (dash-dotted curve),
this structure consists of a central peak that is split in two, along with adjacent Holstein sidebands. For $\mu=230$meV
(solid curve) the split peak reconverges into a single peak, and to a good approximation, the weight under this peak is half that under the split peak that appeared for $\mu_0=97$meV. Increasing $\mu_0$ to $290$meV eliminates entirely this second line. And similar behavior is visible in the next set of peaks at the far right of the figure. So we see the same general progression as in the bare case, despite the richer structure. Again, the behavior is explained in Fig.~\ref{varmu_levels} (lower frame). Focusing again on the peaks around $410$meV, the split peak for $\mu_0=97$meV occurs because the two 
transitions $T_{-2}^1$ and $T_{-1}^2$ no longer have identical energies. The neighbouring Holstein sidebands arise from the transitions $T_{-2_p^0}^1$ and $T_{-1_p^0}^2$. When $\mu_0$ is increased to 230meV, the $T_{-2}^1$ transition (as well as $T_{-2_p^0}^1$) becomes Pauli blocked, leaving a single principle absorption peak arising from the $T_{-1}^2$ transition (and a Holsetin sideband from $T_{-1_p^0}^2$). When $\mu_0$ is increased further, to 290meV, the remaining transitions become Pauli blocked and no peaks in this region appear. The transitions giving rise to the peaks around 540meV in Fig.~\ref{varmu} can similarly be traced.

From such analyses of the level diagrams, we see that the general behavior is maintained in the dressed case because not only are the dressed LLs (to a good approximation) simply shifted down by a factor of  $1/(1+\lam)$ from their bare energies, but any significant split levels or phonon-assisted peaks are clustered close to this energy. This means that the transitions $T_{-n}^{n+1}$ and $T_{-(n+1)}^n$---and any transitions involving associated split levels or phonon-assisted peaks---will be clustered near the energies $(M_{n+1}-M_{-n})/(1+\lam)$ and $(M_{n}-M_{-(n+1)})/(1+\lam)$, respectively, as noted previously. So there remains an identifiable structure composed of one or more nearby peaks, and the weight in this structure decreases by half as groups of transitions (rather than a single transition) become Pauli-blocked. Also, this situation cannot be further complicated by a final state being split in two (or having a strong phonon-assisted peak associated with it) such that the Fermi energy can be moved between the two substituent levels of that final state, because any splitting must occur at least $\simeq\omega_E$ away from the Fermi energy. Furthermore, if the scattering rate were to increase substantially with energy, as has been observed in experiment,\cite{Orlita:11} the differences between the bare and dressed curves would become further obscured. The split peak would appear as a single peak, and the Holstein sidebands would form a smooth background between prominent lines---a larger background than in the bare case, but this difference might be hard to isolate. The effects of changing $\mu$ could also be somewhat obscured by local variations in chemical potential, which are known to occur in graphene due to the existence of electron- or hole-rich puddles.\cite{Martin:08} In such cases, experiments should observe an approximate average of our results for several different values of $\mu$.

\begin{figure}[tb]
\begin{center}
\includegraphics{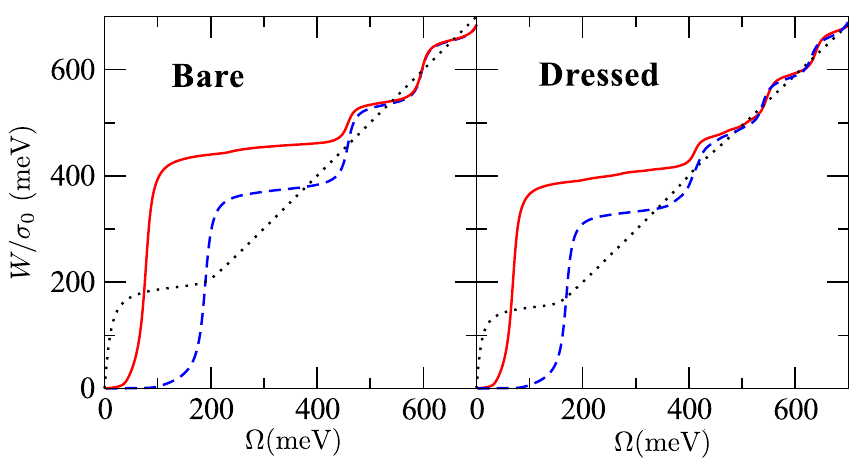}
\end{center}
\caption{(Color online) Partial optical sums for $\mu_0=97$meV (dashed blue curves) and $\mu_0=230$meV (solid red curves), with the same parameters as in Fig.~\ref{varmu}. For comparison, the $B=0$ case with $\mu_0=97$meV (dotted black curves) is also shown. Despite the splitting of levels in the dressed case, the optical weight follows the same pattern as in the bare case. \label{varmu_weight}}
\end{figure}

Another way to visualize the changes in weight with changing chemical potential is to consider the optical weight
up to a cutoff $\Omega$,
\begin{align}
W=\int_0^{\Omega}\sxx(\Omega')d\Omega'.
\end{align}
Results for $W/\sigma_0$ are shown in Fig.~\ref{varmu_weight}
for $\mu_0=97$meV (dashed blue curves) and $\mu_0=230$meV (solid red), with all other parameters as in Fig.~\ref{varmu}. For comparison, we also show the
$B=0$ limit (dotted black curves, studied previously in Ref.~\cite{Carbotte:10}) for $\mu_0=97$meV. In the bare $B=0$ curve, there is a sharp rise in $W(\Omega)$, corresponding to integrating over the Drude peak of half-width $2\Gamma$, followed by a plateau of height $2\mu_0$, followed by a region of linear increase, corresponding to integrating over the constant universal background $\sigma_0$ that sets in at $\Omega=2\mu_0$. The dressed case differs from the bare one by a slightly more rapid initial rise, a lower plateau, and an earlier onset of the linear stage, due to the rescaling of $\Gamma$ and $\mu$ by $1/(1+\lambda)$. Turning now to the finite $B$ curves, we see successive
plateaus arising from integration over successive absorption peaks, leading to oscillation about the $B=0$ curves. The first plateau in each curve arises from an intraband transition, the others from interband transitions. As expected, for $\mu_0=230$meV there is roughly half the weight in the first interband peak as there is for $\mu_0=97$meV, but the total weight is conserved. The heights of the plateaus are reduced in the dressed case relative to their bare values because of the depletion of the quasiparticle lines by the usual factor of $1/(1+\lambda)$.
This missing spectral weight is transferred to the phonon-assisted background,
which causes the plateaus at high energy to appear less sharply defined in the dressed case than in the bare one.


\begin{figure}[tb]
\begin{center}
\includegraphics{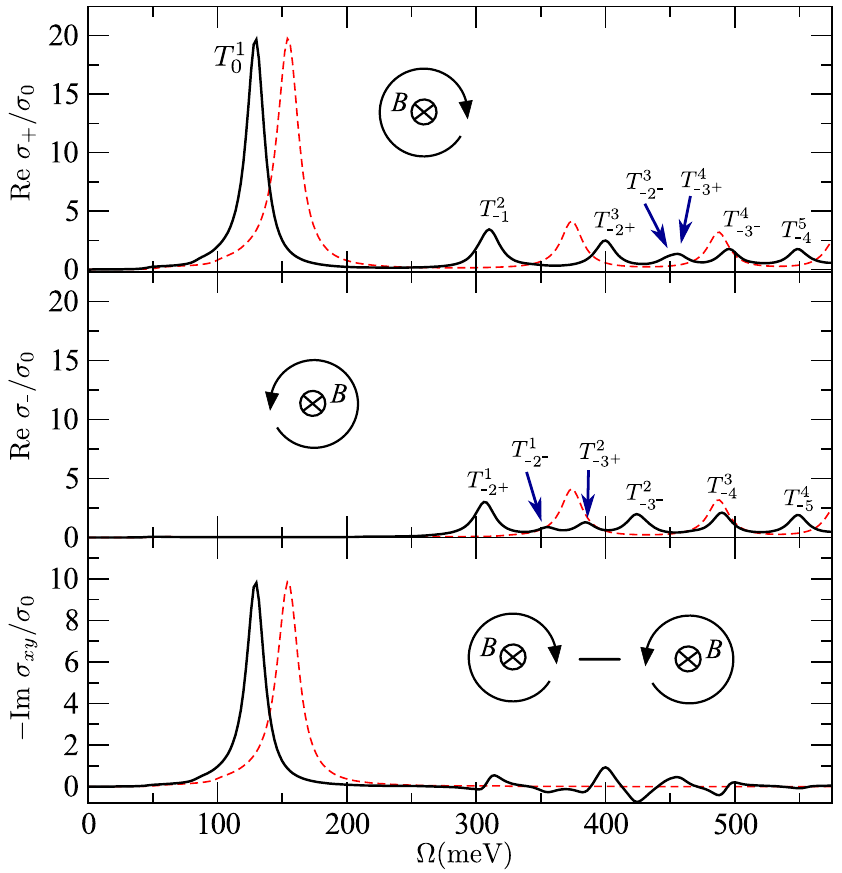}
\end{center}
\caption{(Color online) Bare (dashed red curves) and dressed (solid black curves) optical conductivity as a function of frequency for right- and left-handed circularly polarized light, with parameters $B=18.2$T, $\mu_0=100$meV, $\Gamma=5$meV, $\omega_E=200$meV, and $A=250$meV. The bottom frame shows the imaginary part of the Hall conductivity, equal to one-half the difference of $\sp$ and $\sm$. Additional features not seen in the bare case appear in this difference, due to the renormalized LLs being asymmetrically arranged about $E_0$. \label{asymmetry}}
\end{figure}

\section{Circular Polarization}\label{circular}

So far we have presented results only for the diagonal or longitudinal
conductivity, $\sigma_{xx}(\Omega)$. If the polarization of the incident
light is taken into account, the relevant quantities are $\sigma_{\pm}(\Omega)$,
for right- ($+$) and left-handed ($-$) circularly polarized light,
respectively, and the absorptive part is $\spm(\Omega)$.
Given $\sigma_{\pm}(\Omega)=\sigma_{xx}(\Omega)\pm i\sigma_{xy}(\Omega)$, this absorptive
part is
\begin{align}
\spm(\Omega)=\sxx(\Omega)\mp \sxy(\Omega),
\end{align}
which can be calculated from Eqs.~(\ref{sigma_xx}) and (\ref{sigma_xy}).
Polarized light has important experimental utility, having been used recently
in experiments that have revealed a giant 
Faraday rotation effect in single layer graphene.\cite{Crassee:11a} 
This corresponds to a rotation
of polarized light due to a magnetic field.\cite{Crassee:11b} And as we examine here,
for finite chemical potential $\sp(\Omega)$ and $\sm(\Omega)$ will carry different signatures of
electron-phonon coupling, such that the difference between them provides a direct measure of 
the coupling.

In the bare case, it follows from Eq.~\eqref{sigma_general} that
\begin{align}
&\sp(\Omega) =\frac{2\sigma_0 M_1^2}{\Omega}\sum_{n=0}^\infty\nonumber\\
&\quad \times\left[\delta\left(E_{n+1}-E_{n}-\Omega\right)\theta\left(E_{n+1}\right)\theta\left(-E_{n}\right)\right.\nonumber\\
&\quad + \left.\delta\left(E_{n+1}-E_{-n}-\Omega\right)\theta\left(E_{n+1}\right)\theta\left(-E_{-n}\right)\right]\\
&\sm(\Omega) =\frac{2\sigma_0 M_1^2}{\Omega}\sum_{n=0}^\infty\nonumber\\
&\quad \times\left[\delta\left(E_{-n}-E_{-(n+1)}-\Omega\right)\theta\left(E_{-n}\right)\theta\left(-E_{-(n+1)}\right)\right.\nonumber\\
&\quad + \left.\delta\left(E_{n}-E_{-(n+1)}-\Omega\right)\theta\left(E_{n}\right)\theta\left(-E_{-(n+1)}\right)\right].
\end{align}
 That is, one of the two possible interband and one of the two possible intraband transitions contributes to each $\sp^{(n)}$, and the other two contribute to $\sm^{(n)}$. Hence, if the Fermi energy lies between $E_N$ and $E_{N+1}$ (i.e., $M_N<\mu_0<M_{N+1}$), with $N\geq0$,  we arrive at
\begin{align}
\sp(\Omega) &= \frac{2\sigma_0 M_1^2}{\Omega}\bigg[\delta\left(M_{N+1}-M_{N}-\Omega\right)\nonumber\\
&\quad + \delta\left(M_{N+1}-M_{-N}-\Omega\right)\nonumber\\
&\quad + \sum_{n=N+1}^\infty\delta\left(M_{n+1}-M_{-n}-\Omega\right)\bigg]\label{sp_simplified}\\
\sm(\Omega) &= \frac{2\sigma_0 M_1^2}{\Omega}\sum_{n=N+1}^\infty\delta\left(M_{n}-M_{-(n+1)}-\Omega\right).
\end{align}
Because the bare levels are arranged symmetrically about the Dirac point---i.e., $M_{-n}=-M_n$---the terms in the two sums are identical. Therefore, in the bare case $\sp$ and $\sm$ have an identical set of peaks, except that $\sp$ has an intraband peak and one more interband peak than $\sm$ (with the situation reversed if $N<0$). But this relies on the symmetric arrangement of levels, which no longer exists in the dressed case. And so we expect that in the dressed case, these sets of peaks will differ.

This is illustrated in Fig.~\ref{asymmetry}, which shows $\sp(\Omega)$ (top frame), $\sm(\Omega)$ (middle), and the difference $[\sp(\Omega)-\sm(\Omega)]/2=-\sxy(\Omega)$ (the imaginary part of the off-diagonal or transverse Hall conductivity). The bare case is shown in dashed red; the dressed, in solid black. Here we use $\mu_0=100$meV and $B=18.2$T, such that the Fermi energy lies between the zeroth and first LLs. With this arrangement, the lowest-frequency interband transition is degenerate with the intraband one, such that $\sp(\Omega)$ has only one low-frequency peak that is absent from $\sm(\Omega)$. In the bare case, that single peak in $\sp(\Omega)$ is followed by a series of interband peaks that are identical to those in $\sm(\Omega)$. But as expected, in the dressed case the interband peaks in $\sp(\Omega)$ differ significantly from those in $\sm(\Omega)$.  If $\sp(\Omega)$ and $\sm(\Omega)$ are summed 
and the sum is divided by 2, we obtain $\sigma_{xx}(\Omega)$. In the bare case, this has peaks at the same positions as those in $\sp(\Omega)$ but with the first peak halved in size. In the dressed case, the sum of the two sets of interband peaks becomes complicated. On the other hand, if the difference is taken and then halved to obtain $-\sxy(\Omega)$, then in the bare case all the interband peaks cancel, as shown in the bottom frame of Fig.~\ref{asymmetry}. In the dressed case, there are instead oscillations about zero above 300meV.

\begin{figure}[tb]
\begin{center}
\includegraphics{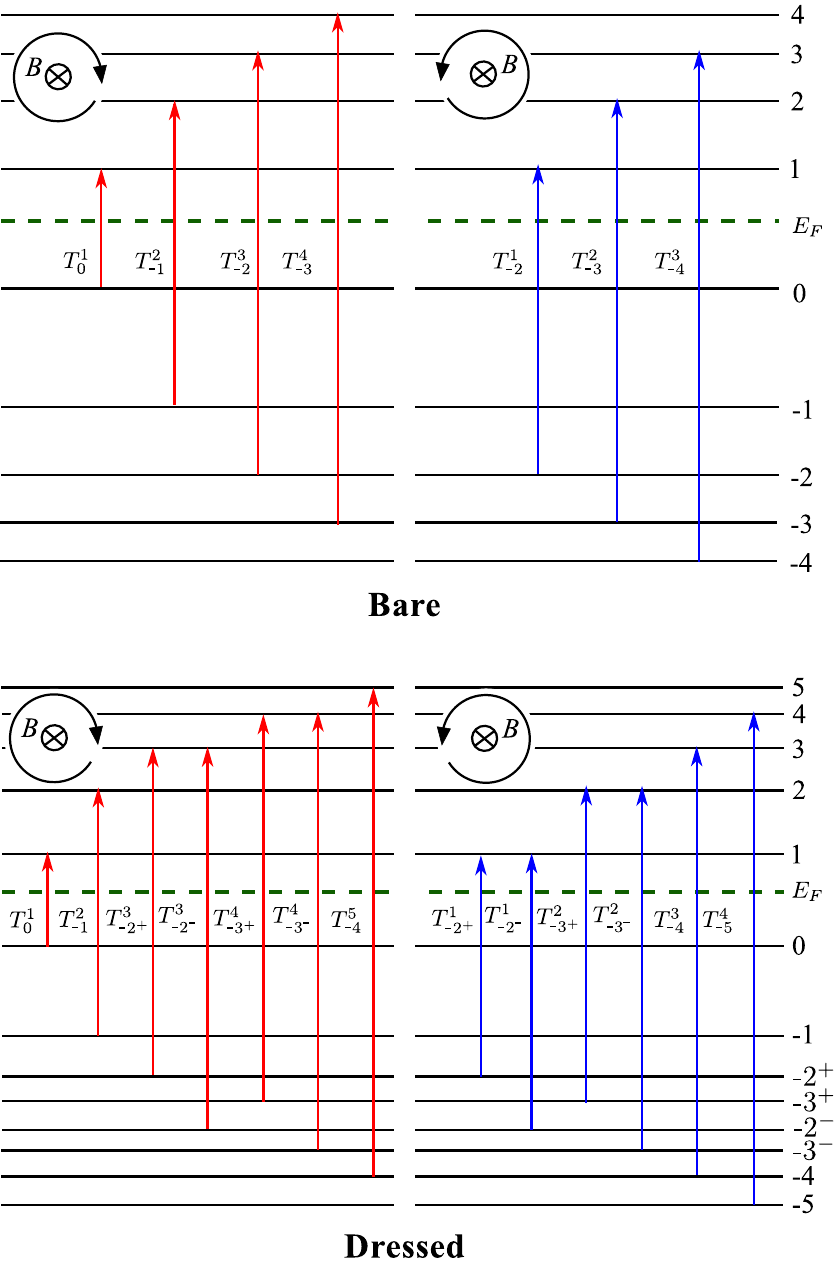}
\end{center}
\caption{(Color online) Level diagrams corresponding to the conductivities shown in Fig.~\ref{asymmetry}. In the dressed case, level spacings are not symmetric about the zeroth level; most significantly, the $n=-2$ and $-3$ LLs are split, while the $n=2$ and 3 are not. This asymmetry causes peaks in $\sp$ and $\sm$ to appear at differing energies, as seen in Fig.~\ref{asymmetry}. \label{asymmetry_levels}}
\end{figure}

Figure~\ref{asymmetry_levels} shows the level diagram and transitions giving rise to the various peaks, and we can see therein that the largest difference between the levels above the Dirac point and those below is the splitting of the $-2$ and $-3$ LLs. This shows up as the largest oscillation in the lower frame of Fig.~\ref{asymmetry}. However, even in the absence of splitting, oscillations are apparent. And since the bare case has no such oscillations, any nonzero result in this region is a direct consequence of correlation effects and could be used to estimate their magnitude.

\section{Semiclassical cyclotron resonance}\label{cyclotron}

The semiclassical limit comes about when the quantization associated with the Landau levels no longer plays a prominent role. This corresponds to a large chemical potential, $\mu\gg M_1$, such that $\mu$ falls between levels $N$ and $N+1$ where $N\gg1$. For simplicity, we take $\mu>0$ in this discussion. Note that if $\mu_0$ falls between the bare levels $N$ and $N+1$, then $\mu$ falls between the same two dressed levels, and so we can freely refer to the relative position of either $\mu_0$ or $\mu$. This is the case because in this regime, $\mu\simeq \mu_0/(1+\lam)$ and the two levels fall very near the Fermi energy (and hence very far from any peaks in the self-energy), meaning they, too, are simply shifted down by $1/(1+\lam)$. Now, because $N$ is large, the energy of an intraband transition between these two levels is given by $\delta M\equiv M_{N+1}-M_N\simeq \sqrt{\frac{eBv_F^2\hbar}{2Nc}}$. Noting that the chemical potential $\mu_0$ is approximately equal to $M_N$, we find
\begin{equation}
\delta M\simeq\frac{v_F^2 eB\hbar}{\mu_0c},\label{deltaM}
\end{equation}
which is the semiclassical cyclotron resonance (CR) frequency, $\omega_{cr}$. In this limit, the level spacing at the Fermi energy, and therefore the transition energy, is linear in the magnetic field. Such linear dependence, which has been measured in a recent experiment,\cite{Witowski:10} sharply contrasts with the behavior for $\mu\lesssim M_1$, where the spacing and transitions vary as $\sqrt B$.

\begin{figure}[tb]
\begin{center}
\includegraphics{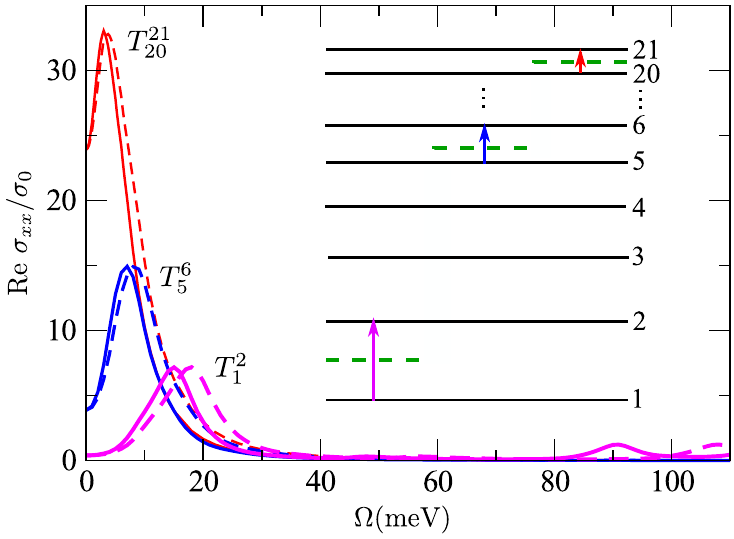}
\end{center}
\caption{(Color online) Optical conductivity corresponding to cyclotron resonance for three values of chemical potential: $\mu_0=202.5$meV (top two curves), $\mu_0=105$meV (middle two), and $\mu_0=55$meV (bottom).  The remaining parameters are $B=1.5$T, $\Gamma=2.5$meV, $\omega_E=200$meV, and $A=250$meV. Dashed curves are bare, and solid are dressed. The inset shows the level diagram and transitions giving rise to the absorption peaks, with the thick, dashed, green curve marking the location of the Fermi level for the three values of chemical potential. At the lower right, we see the transition from semiclassical to quantum behavior as $\mu$ decreases and more peaks (due to $-3\to2$ and $-2\to3$ transitions) become visible at reasonably low energies.\label{CR}}
\end{figure}

We illustrate the transition between the semiclassical and quantum regimes in Fig.~\ref{CR}, which shows $\sxx/\sigma_0$ versus $\Omega$ for three values of chemical potential, with solid curves for the dressed case and dashed for the bare case. Here we have used a relatively small magnetic field, $B=1.5$T, corresponding to $M_1=45$meV, to make the transition between regimes visible within a relatively small energy range. The top two curves (in red) are for $\mu_0=202.5$meV$\gg M_1$, deep in the semiclassical regime; the middle two (in blue), for $\mu_0=105$meV; and the bottom two (in magenta), for $\mu_0=55$meV$\sim M_1$, in the quantum regime. In these curves, the peaks at low frequency represent the cyclotron resonance. Along with the conductivity, we show the level diagram and the transitions responsible for those low-frequency peaks, with the Fermi energy indicated by the thick, dashed, green line. As is obvious from the diagram, when $\mu$ is increased the energy of the intraband transition is decreased while the energy of the interband transitions are increased. The onset of the quantum regime coincides with the interband transitions becoming relevant on the scale of the CR, which can be seen in the bottom set of curves, where an additional set of peaks due to interband transitions become visible at higher energies.

\begin{figure}[tb]
\begin{center}
\includegraphics{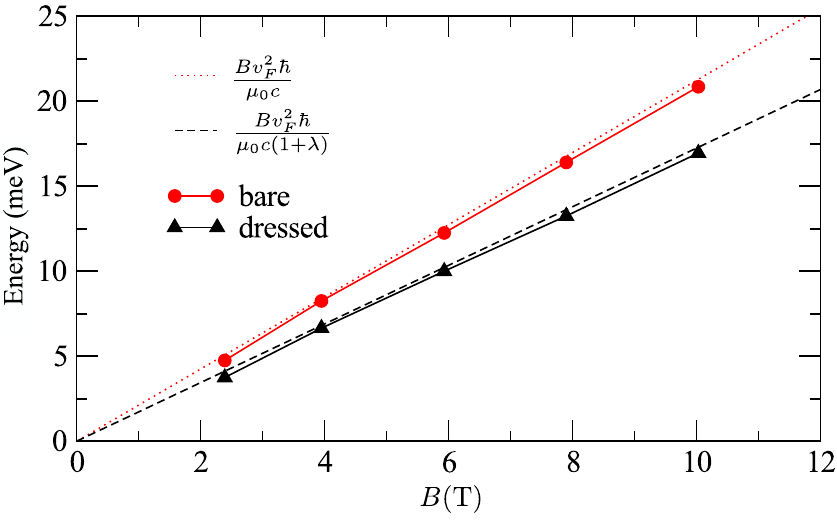}
\end{center}
\caption{(Color online) The cyclotron resonance frequency $\omega_{cr}$ as a function of $B$ for the same parameter values as in Fig.~\ref{CR_peaks}. In both the bare and dressed cases, $\omega_{cr}$ closely tracks the expected linear approximation, represented by the dotted red curve in the bare case and the dashed black curve in the dressed case. \label{wc(B)}}
\end{figure}

Note that all of this holds equally well in both bare and dressed cases; as can be seen in the figure, the dressed conductivity differs from the bare conductivity mainly by a downward shift in frequency. Because the relevant levels are very near the Fermi energy, the cyclotron resonance is unaffected by the split levels and phonon-assisted peaks discussed in the preceding sections. This allows us to easily gain analytical understanding of many features of the optics and of renormalization in this regime. We first note that one can perform the same calculation that led to Eq.~\eqref{deltaM} on the dressed energies $E_n=\Sigma_1(E_n)-\mu+M_n$. Using $E_n\simeq\frac{M_n-\mu_0}{1+\lambda}$, one finds the renormalized cyclotron frequency
\begin{equation}
\omega_{cr}\simeq \frac{v_F^2eB\hbar}{c\mu_0(1+\lambda)},\label{wc}
\end{equation}
which is simply the bare frequency $\omega_{cr} = v_F^2 \frac{eB\hbar}{c\mu_0}$ divided by $1+\lambda$. Figure~\ref{wc(B)} confirms this approximate result, showing the approximations \eqref{deltaM} and \eqref{wc} together with the cyclotron frequency as determined from full numerical results (indicated by circles in the bare case and triangles in the dressed case). Note that because the dressed chemical potential $\mu$ is approximately $\mu_0/(1+\lambda)$, the bare formula for $\omega_{cr}$ remains valid so long as bare quantities are replaced by renormalized ones: $\omega_{cr}=\frac{v_F^2 B\hbar}{c\mu}$, where $v_F$ now refers to the renormalized Fermi velocity.

\begin{figure}[tb]
\begin{center}
\includegraphics{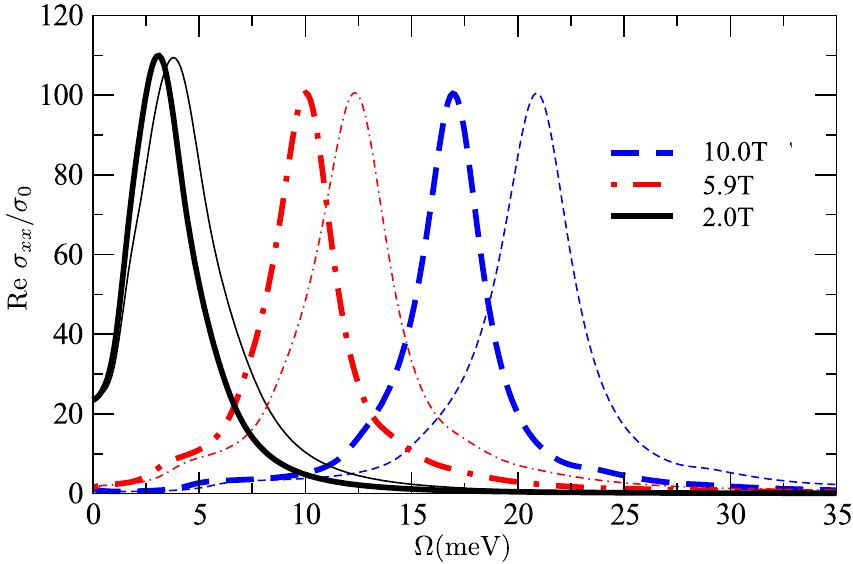}
\end{center}
\caption{(Color online) Bare (thin curves) and dressed (thick curves) conductivity corresponding to cyclotron resonance for three values of magnetic field. The cyclotron resonance frequency increases with increasing $B$, as the level spacing at the Fermi energy increases. The remaining parameters are $\mu_0=310$meV, $\Gamma=1.0$meV, $\omega_E=200$meV, $A=250$meV.\label{CR_peaks}}
\end{figure}

Figure~\ref{CR_peaks} more fully illustrates the effects of renormalization on the CR peaks. The diagonal conductivity as a function of $\Omega$ is shown in the bare (thin curves) and dressed (thick curves) cases for three values of $B$ at a fixed, large value of the chemical potential, $\mu_0=310$meV. Increasing $B$ from 2.0T (solid black curves) to 5.9T (dot-dashed red) to 10.0T (dashed blue) causes the CR peak to move to higher energies, as discussed above. And as we saw in the preceding sections, the width of the peak in each case is decreased due to renormalization. The height of the peaks, however, is only weakly affected by the varying field strength and almost entirely unaffected by renormalization. All of these features follow from the fact that the relevant levels are near the Fermi energy, far from any phonon-assisted peaks. This means Eq.~\eqref{dressed_peak} is valid. Replacing $n$ with $N$ therein and substituting Eq.~\eqref{deltaM} into it, we find
\begin{equation}
\sxx(\Omega) = \sigma_0\frac{2\mu_0}{1+\lambda}\frac{1}{\pi}\frac{2\Gamma/(1+\lambda)}{(\Omega-\omega_{cr})^2+(\frac{2\Gamma}{1+\lambda})^2},
\end{equation}
where $\omega_{cr}$ here stands for $(M_{N+1}-M_N)/(1+\lambda)$, which is approximately the renormalized cyclotron resonance frequency given in Eq.~\eqref{wc}. Here we see the width of the dressed CR peak is reduced by a factor of $1/(1+\lambda)$. We also see that the value of $\sxx$ at $\omega_{cr}$,
\begin{equation}
\sxx(\Omega=\omega_{cr}) = \sigma_0\frac{\mu_0}{\pi\Gamma},
\end{equation}
is independent of the magnetic field strength and of the electron-phonon coupling, as we saw in the full numerical results of Fig.~\ref{CR_peaks}. This contrasts with the behavior in the quantum regime, where the height is unaffected by coupling but varies as $\sqrt{B}$, as we saw in Eq.~\eqref{dressed_height}. Note that as with $\omega_{cr}$, when $\mu \simeq\mu_0/(1+\lambda)$, we can write this result in terms of observable, dressed quantities: ${\rm Re}\sigma_{xx}(\Omega=\omega_{cr}) = \sigma_0\frac{\mu}{\pi\Gamma}$, where $\Gamma$ is here the dressed width of the spectral peaks.

As we can see from the $B=2.0$T curves in Fig.~\ref{CR_peaks}, this prediction of $B$-independent peak amplitude begins to fail at small $B$, for which $\omega_{cr}$ gets close to zero. The failure is due to the effects of finite width, which allow neighbouring transitions, rather than the $T_N^{N+1}$ transition alone, to contribute significant optical weight in the energy range of the cyclotron resonance.

\begin{figure}[tb]
\begin{center}
\includegraphics{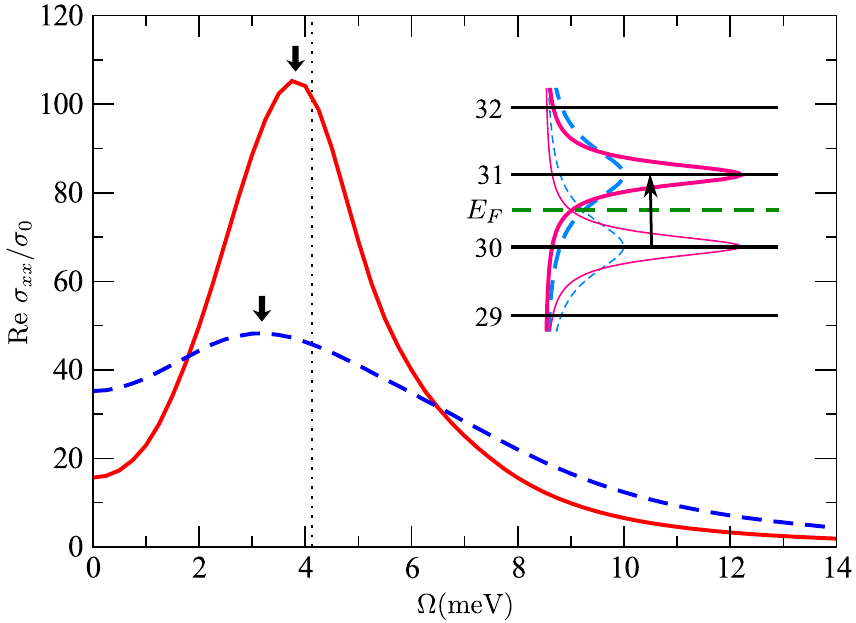}
\end{center}
\caption{(Color online) Effect of finite broadening on the cyclotron resonance frequency for $B=2.4$T, $\mu_0=310$meV, $\omega_E=200$meV, and $A=250$meV. The solid red curve is for $\Gamma=1.0$meV; the dashed blue curve, for $\Gamma=2.5$meV. Black arrows point to the peak position in each case, which shifts away from the expected value $Bv_F^2/[\mu_0(1+\lambda)]$ (marked by the vertical dotted line) as $\Gamma$ increases. The inset shows the levels and transition between them giving rise to the cyclotron resonance, with the widths of the levels causing noticeable portions of the initial (or final) state to fall above (or below) the Fermi energy.\label{Gamma_effect}}
\end{figure}

The finite width of the levels also causes a shift in the value of the cyclotron resonance frequency. Figure~\ref{Gamma_effect} compares the conductivity for two values of $\Gamma$: 2.5meV (dashed curve) and 1meV (solid curve). The dotted vertical line indicates the value of $\omega_{cr}$ expected from Eq.~\eqref{wc}, while the two black arrows point to the peak positions in the two cases. For both values of $\Gamma$, the peak positions differ from the expected result, with a larger difference for the larger value of $\Gamma$. This effect of $\Gamma$ can be understood from the inset, which shows the levels and transition between them that gives rise to the cyclotron resonance peak. Significant portions of the initial (final) level fall above (below) the Fermi energy, causing the average transition energy to deviate from the difference $E_{31}-E_{30}$. Results here include coupling to a phonon, but obviously this effect occurs in the bare case as well.

\section{Effects of broadening and comparison with experiment}\label{experiment}
Several experimental studies of optical conductivity in graphene in magnetic field have been performed in both ultrathin epitaxial graphite samples\cite{Sadowski:06,Sadowski:07,Orlita:10,Orlita:11} and in single-layer graphene.\cite{Jiang:07,Henriksen:10,Deacon:07} These experiments have not seen obvious evidence of the more noteworthy features we have described, involving highly deformed or split peaks. There are two reasons for this. First, typically the experiments observe only the lowest few absorption peaks, arising from transitions between low LLs. Since signatures of electron-phonon coupling occur when the initial or final level in a transition has an energy of magnitude $\gtrsim\omega_E$, this means that one requires fairly large magnetic fields to see such signatures (given that $\omega_E$ is typically near $150$meV or $200$meV in experiments). Second, in order to see fine details of the spectrum, such as split peaks, one requires fairly small scattering rates. In the experiments with large magnetic fields, the scattering rates have been significantly larger than the values we have used.\cite{Jiang:07,Henriksen:10} In experiments with smaller magnetic fields, there is an indication that the scattering rate increases linearly with energy,\cite{Orlita:11} such that peaks are obscured at the large energies where signatures of coupling would be seen. In cases such as these, in which the scattering rate is comparable to or greater than the energy difference between split lines, double peaks or several nearby peaks will merge. This changes the signature of coupling from a split peak to a broadened peak.

\begin{figure}[tb]
\begin{center}
\includegraphics[width=\columnwidth]{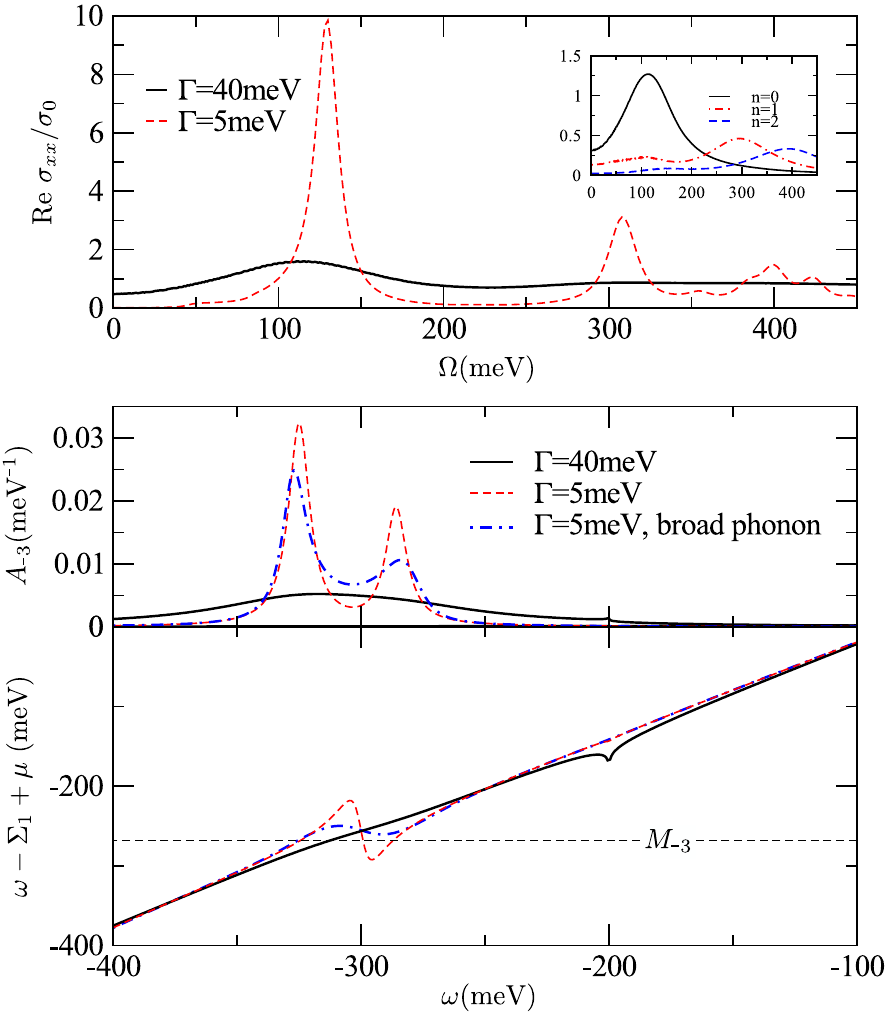}
\end{center}
\caption{(Color online) Effects of large broadening on the spectrum. The upper frame shows the diagonal conductivity for $\Gamma=40$meV and $5$meV; the inset shows how the separate contributions to $\sxx$ overlap for the $\Gamma=40$meV case. The lower two frames show a spectral function, $A_{-3}$, involving a split peak, along with the function $\omega-\Sigma_1+\mu$ that determines the nature of the peaks. Here results are shown for both a large impurity scattering and a broadened phonon; either effect, if sufficiently large, can annul the splitting of levels. In all cases, $B=18.2$T, $\mu_0=100$meV, $\omega_E=200$meV, $A=250$meV. The broadened phonon distribution is a truncated Lorentzian with parameters $A'=555$meV, $\delta=15$meV, and $\delta_0=30$meV.\label{broadening}}
\end{figure}

The effect of large broadening is illustrated in Fig.~\ref{broadening}. In the upper frame,
we show results for $\sxx(\Omega)$ with $\Gamma=5$meV (dashed
red curve) and $\Gamma=40$meV (solid black). For the larger value of $\Gamma$, little remains of the rich structure seen in the dashed curve.
In the inset, we show the $\Gamma=40$meV results decomposed into $\sxx^{(0)}$, $\sxx^{(1)}$, and $\sxx^{(2)}$, allowing us to see how strongly the peaks overlap.
In the lower two frames, we show results for $A_{-3}(\omega)$ (middle) and 
$\omega-\Sigma_1(\omega)+\mu$ (bottom) for both of the above two cases and for the case where $\Gamma$ remains small (5meV) but the phonon spectrum is broadened 
(dash-dotted blue curve). In this case, we model
the electron-phonon spectral density $\alpha^2F(\nu)$ of Eq.~(\ref{Sigma})
with the truncated Lorentzian form used in Ref.~\cite{Carbotte:10}, namely,
\begin{align}
\alpha^2F(\nu)=\frac{A'}{\pi}\biggl[\frac{\delta}{(\nu-\omega_E)^2+\delta^2}-
\frac{\delta}{\delta_0^2+\delta^2}\biggr]\theta(\delta_0-|\omega_E-\nu|),
\label{Pnu}
\end{align}
which is peaked at $\omega_E$ with half-width $\delta$ and truncated at energy $\omega_E\pm\delta_0$. The energy $A'$ is adjusted to give the same value of $\lambda$ obtained from the Einstein spectrum. Comparing the three curves, we see that for the features around $-300$meV, broadening the phonon has the same effect as increasing the residual scattering, reducing the oscillation in the self-energy (seen in the lower frame) and hence tending to merge the peaks in the spectral function (in the middle frame). For the very large scattering rate $\Gamma=40$meV, this oscillation has entirely disappeared. To make the three curves clearly distinguishable, we have chosen the width of the phonon spectrum such that the behavior of the dash-dotted blue curves lie midway between that of the solid black and dashed red curves. The feature at $\omega=-200$meV, visible only in the $\Gamma=40$meV case, is the logarithmic singularity in the self-energy (for comparison, see the inset of Fig.~\ref{A0} and middle frame of Fig.~\ref{A0A1}). For finite chemical potential, as in these curves (where $\mu_0=100$meV), this logarithmic singularity is generally small but grows roughly linearly with $\Gamma$,\cite{Pound:11b} making it noticeable when $\Gamma=40$meV; on the other hand, it is weakened by broadening the phonon spectrum.

\begin{figure}[tb]
\begin{center}
\includegraphics[width=\columnwidth]{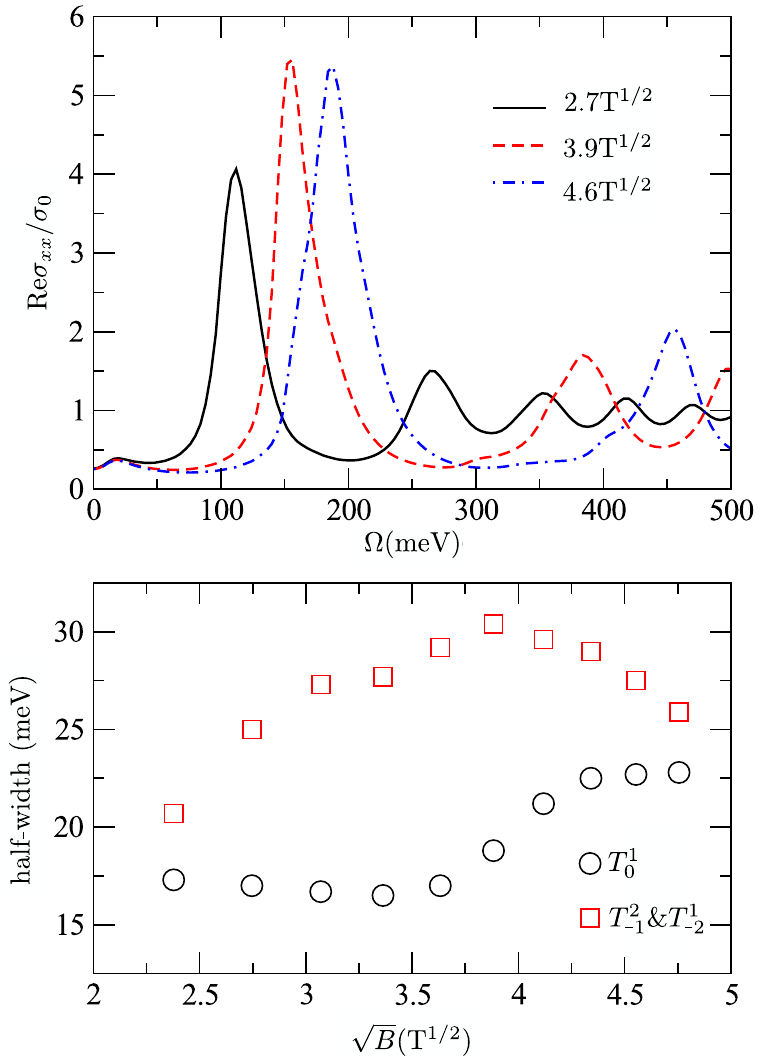}
\end{center}
\caption{(Color online) Effect of varying magnetic field on the conductivity for $\Gamma=10$meV, $\mu_0=15$meV, $\omega_E=150$meV, $A=150$meV, and $v_F=1.27\times10^6$m/s. The upper frame shows the diagonal conductivity as a function of photon frequency for three different values of $B$; to allow easy comparison with the lower frame, the legend indicates the values of $\sqrt{B}$ for which each curve was calculated. The lower frame shows the half-widths of the first two peaks in the conductivity as a function of $\sqrt{B}$. Circles indicate the half-width of the first peak, which arises from the transition $T^1_0$, while squares indicate that of the second peak, which arises from the two transitions $T_{-2}^1$ and $T^2_{-1}$. \label{Jiang}}
\end{figure}

In cases where split peaks become merged in this manner, one can detect electron-phonon coupling via broadened, rather than split, peaks. Specifically, we can expect a jump in the broadening of an absorption peak when the initial or final LL in the transition lies near the phonon energy. Such a jump is displayed in Fig.~\ref{Jiang}. The upper frame shows conductivity curves for three values of $\sqrt{B}$, while the lower frame shows the half-widths of the first two absorption peaks as a function of $\sqrt{B}$. In the lower frame, the half-width of the first absorption peak (open circles) jumps at around $16$T. For the chemical potential used here ($\mu_0=15$meV), the first absorption peak is created by the $0\to1$ transition, and the jump occurs when the $n=1$ LL passes through the phonon frequency. Physically, this corresponds to the opening of another decay channel for the electron via the creation of a phonon. A similar jump has been seen in the data of Jiang et al.,\cite{Jiang:07,Henriksen:10} and to generate the data of Fig.~\ref{Jiang}, we have chosen parameters similar to theirs: specifically, a somewhat larger scattering rate than used through most of the paper, with $\Gamma=10$meV, and a larger Fermi velocity, with $v_F=1.27\times10^6$meV (which agrees with their inferred value once renormalized by $1/(1+\lambda)$). The jump they see occurs at $\sqrt{B}\simeq3.5$T$^{1/2}$, and to obtain a jump at a similar field strength, we have chosen $\omega_E=150$meV (a typical phonon energy reported in some experiments\cite{Bianchi:10,Li-G:09,Brar:10}). However, a precise comparison cannot be made, because Jiang et al. report only the filling factor, rather than chemical potential, for which their results were obtained. The dependence of the jump on chemical potential is easily estimated: As stated above, the jump in half-width occurs when the $n=1$ LL, which lies approximately at $(M_1-\mu_0)/(1+\lambda)$, reaches $\omega_E$. This leads to the condition $(M_1-\mu_0)/(1+\lambda)=w_E$. Rearranging for $\sqrt{B}$, one finds (in Gaussian units)
\begin{equation}
\sqrt{B} = \frac{\omega_E(1+\lambda)+\mu_0}{v_F\sqrt{2e\hbar/c}},
\end{equation}
which shows that the field strength at which the jump occurs depends linearly on $\mu_0$. If the jump seen by Jiang et al. does not vary under changes of chemical potential (in a range that keeps the filling factor roughly constant), then our results cannot plausibly explain it. However, a jump that does depend on chemical potential would be a strong signature of electron-phonon coupling. The resonance frequency at which the jump occurs, $E_1-E_0=\omega_E+\mu$, is also an indicator of whether or not it is caused by coupling to a single phonon.

Figure~\ref{Jiang} also shows, in the upper frame, that the dressed absorption peaks deviate from Lorentzian shape, in addition to their jump in broadening, when the field strength is near or above that at which the broadening occurs. Such asymmetry can be understood from Eq.~\eqref{dressed_peak} and Fig.~\eqref{sigma_n=2}. At low fields, before the $n=1$ LL approaches $\omega_E$, the dressed peak is described by Eq.~\eqref{dressed_peak}, with a Lorentzian shape. Once the $n=1$ level approaches $\omega_E$, transitions made possible by phonon-assisted processes begin to have significant optical weight. As shown in Fig.~\ref{sigma_n=2} (along with neighbouring figures), these transitions are not symmetrically weighted about a central peak. When the broadening is large and the peaks merge, the result will be an asymmetric lineshape. As with the jump in broadening, asymmetric lineshapes have been observed by Jiang et al.\cite{Jiang:07,Henriksen:10} While that asymmetry has been attributed to substrate effects,\cite{Henriksen:10} electron-phonon coupling may contribute.

Returning to the bottom frame of Fig.~\ref{Jiang}, we see that the lifetime for the second absorption peak (open squares) exhibits a quasilinear increase with $\sqrt{B}$, followed by a decrease, with the maximum occurring at approximately the same field strength as did the jump in width of the first peak. The increase brings the width above that of the first peak. Such quasilinear behavior has been observed for lower fields of $\simeq1.5$--2.5T$^{1/2}$ in the data of Orlita et al.\cite{Orlita:11} Unfortunately, data for the second peak's width has not been reported at higher fields, and a turnaround of the width has not been observed at present.

Experiments have also found anomalies in the positions of peaks. Broadly speaking, they have all confirmed the expected $\tilde v_F\sqrt{B}$ dependence of the resonance energies. But while Sadowski et al.\cite{Sadowski:06,Sadowski:07,Orlita:10,Orlita:11} find a common renormalized Fermi velocity $\tilde v_F$ fits all the data, Jiang et al.\cite{Jiang:07, Henriksen:10} find a different velocity is required to fit the two measured level resonances, and Deacon et al.\cite{Deacon:07} find a particle-hole asymmetry in the fitted velocity, which is not seen by the other two groups. Moreover, the different groups deduce markedly different values of $\tilde v_F$, even when fitting the same resonance energies. Our results suggest that these discrepancies cannot be explained by coupling to a phonon of energy in the typical range of 150--230meV. In our calculations, we find that in a large part of the spectrum, $v_F$ is simply renormalized by a factor of $1/(1+\lambda)$ for all resonance energies. The resonance energies deviate from the behavior $\frac{v_F\sqrt{B}}{1+\lambda}$ only when the initial or final level in the relevant transition lies near a phonon-assisted peak, as shown in Fig.~\ref{transition_energies} for the $\mu=0$ case.

In experiment, other issues will complicate the effects of electron-phonon coupling. For example, the detailed nature of a given defect-structure can introduce important energy-dependence into the scattering rate,\cite{Venezuela:11} and finite-size effects can introduce modifications.\cite{Libisch:10,Romanovsky:11} Coupling to acoustic phonons may also occur. As discussed in Ref.~\cite{Pound:11b}, this has qualitatively the same effect as coupling to a single Einstein mode, introducing phonon-assisted peaks in the self-energy at frequencies shifted from the LLs by the Debye frequency $\omega_D$. To be consistent with experiment, the coupling to such phonons must be weak, and since $\omega_D$ will typically be much smaller than the the LL spacing, the result will be small shoulders on the peaks in the self-energy. This, in turn, will lead to small shoulders in the absorption peaks.


\section{Summary and conclusion}\label{summary}

In a conventional metal, coupling of the charge carriers to a phonon provides a transfer of optical spectral weight from the Drude-like coherent peak at zero energy, characteristic of a metallic state, to incoherent boson-assisted Holstein sidebands. In such absorption, a quasiparticle particle-hole pair is created as well as a phonon. We have studied how similar processes alter the optical spectrum in graphene under an external magnetic field $B$, a regime in which the charge carrier bands are discretized into Landau levels.

Our study rests on the selection rules for the allowed optical transitions, which were found to be straightforward generalizations of those in the bare band case. If only bare LLs are considered, the optical absorption spectrum consists of a sequence of peaks arising due to transitions between the levels. In an interband transition, a particle can be excited from a level labeled by an integer $n$ to a level $n+1$; in an interband transition, a particle can be excited from a level labeled by a negative integer $-n$ (positive or zero $n$) to a level $n\pm1$. In addition, the initial state must be occupied (that is, $\omega\leq0$) and the final state unoccupied (that is, $\omega\geq0$). When the charge carriers are coupled to a phonon, the spectral weight in the $n$th level, denoted by $A_n(\omega)$, is redistributed from a single Dirac or Lorentz distribution into multiple peaks---but the selection rules remain essentially the same: a transition from an occupied initial state of weight $A_{n_i}(\omega)$ ($\omega\leq0$) to a final state of weight $A_{n_f}(\omega')$ ($\omega'\geq0$) is allowed if and only if $n_f=n_i+1$ (for an intraband transition) or $n_f=|n_i|\pm1$ (an interband transition). The changes in the optical spectrum can therefore be straightforwardly understood from the changes in the distribution of spectral weight in each level.

Those changes are largely encapsulated by a shift of each level from the bare energy $M_n={\rm sgn}(n)v_F\sqrt{2|n|eB\hbar/c}$ to the lower energy $M_n/(1+\lambda)$, where $\lambda$ is the electron-phonon effective mass renormalization parameter. This can can be interpreted as a renormalization of the Fermi velocity to $v_F/(1+\lambda)$. Correspondingly, the widths of the resulting quasiparticle spectral lines are decreased by a factor of $1/(1+\lambda)$, while their amplitude remains unchanged, leading to a loss of spectral weight in each level. The lost weight is transferred to phonon-assisted peaks. Since $\lambda\lesssim0.2$ in graphene, these have only a small spectral weight. In addition to the shift of spectral weight to phonon-assisted peaks, when a Landau level falls near one of those peaks, it deviates from the simple renormalization of the Fermi velocity, and it can split into two lines sharing the original line's spectral weight. Favorable conditions for this splitting to occur are small broadening and a sharply peaked phonon distribution; if the intrinsic broadening is comparable to the separation of levels, the splitting can instead appear as a single widened peak. Note that while we have formulated a mathematically sharp distinction between a split quasiparticle level and a quasiparticle level with a phonon-assisted sideband, when the spectral weight in one of the substituents of a split level becomes smaller than $\lambda/(1+\lambda)$, it is no longer possible to distinguish between these two cases from an empirical point of view.

Despite these complicated features in each $A_n$, all significant additional absorption peaks tend to fall very near what would be expected from a simple renormalization of the Fermi velocity, with the peaks falling closest to that prediction having the most optical weight. This is because additional peaks of significant weight in each $A_n$ likewise cluster near the energy expected from the simple renormalization of $v_F$. It follows that many features in the optics in the bare case, such as the change of optical weight under an absorption peak as the chemical potential is varied, largely carry over into the dressed case, but with simple Lorentzians replaced by deformed and slightly split absorption peaks generated from multiple slightly different transitions.

As a result of the redistribution of spectral weight, the renormalized states are no longer symmetrically arranged about the zeroth LL. This asymmetry implies that the peaks in the real parts of the conductivity in the right- ($\sigma_+$) and left-handed ($\sigma_-$) circular bases no longer align in energy as they would in the bare case. Consequently, the imaginary part of the transverse conductivity, given by the difference $\sxy=-\frac{1}{2}(\sp-\sm)$, now possesses additional peaks and changes of sign, though these features will likely be small in practice.

We have also considered the semiclassical limit, which corresponds to the chemical potential becoming large compared to energy scale of the Landau levels: $\mu\gg M_1$. In this limit, the levels near the Fermi energy are densely spaced. This means that the transition giving rise to the first peak in $\sxx$, at the cyclotron resonance frequency $\omega_{cr}$, is between levels very near the Fermi energy, and is hence unaffected by the more complicated features that appear in the spectrum at frequencies above $\omega_E$. We find that $\omega_{cr}$ is thus given by a simple renormalization of the bare-band approximate formula $\omega_{cr} = Bv_F^2/\mu_0$ by a factor of $1/(1+\lambda)$. If we take into account that the dressed chemical potential $\mu$ is approximately $\mu_0/(1+\lambda)$, we can write $\omega_{cr}$ simply in terms of renormalized quantities as $B v_F^2/\mu$, where $v_F$ is the renormalized Fermi velocity. We find, however, that because the spacing is small, the finite width of the levels slightly decreases the cyclotron frequency from the expected value.

Lastly, we have discussed the relevance of our results to current experiments. In the case of large broadening, we have shown that electron-phonon coupling leads to a signature jump in the half-width of absorption peaks as LLs are made to pass through the phonon frequency. Such jumps may have been seen in the results of Jiang et al.,\cite{Jiang:07,Henriksen:10} and they should be readily identifiable based on their known dependence on parameters such as chemical potential.


\begin{acknowledgments}
We thank Erik Henriksen and Marek Potemski for helpful discussions and the anonymous referees for suggested improvements. This research was supported in part by the National Science Foundation under Grant No. NSF PHY05-51164, the Canadian Institute for Advanced Research, and the Natural Sciences and Engineering Research Council of Canada.
\end{acknowledgments}

\appendix

\section{Selection rules}\label{selection_rules}
From the earlier work of Gusynin et al.,~\cite{Gusynin:07a} the real part of the conductivity in the circular basis is given by
\begin{align}
\spm(\Omega) &= \sigma_0\frac{M_1^2}{\Omega}\sum_{n=0}^\infty\int_0^\Omega d\omega\left[(1\mp1)\psi_{n,n+1}(\omega,\omega-\Omega)\right.\nonumber\\
&\quad +(1\pm1)\left.\psi_{n,n+1}(\omega-\Omega,\omega)\right].\label{sigma_pm}
\end{align}
Following the discussion in Sec.~\ref{spectral_functions}, we can model each $A_n(\omega)$ as a set of lines, one at each renormalized quasiparticle level $n^\alpha$ and one at each phonon-assisted peak $n_p^m$, as in Eq.~\eqref{A_n model}. We reproduce that equation here for convenience:
\begin{equation}
A_n(\omega) = \sum_\alpha W_n^\alpha\delta(\omega-E^\alpha_n)+\sum_{m=-\infty}^\infty W_{m,n}\delta(\omega-P_{m,n}).
\end{equation}
Substituting this into Eq.~\eqref{sigma_pm} and simplifying the resultant Heaviside functions, we arrive at
\begin{widetext}
\begin{align}
\spm(\Omega) &=\sigma_0\frac{M_1^2}{\Omega}\sum_{n=0}^\infty\Bigg\lbrace
(1\mp1)\bigg[\sum_{\alpha,\beta}W_n^\alpha W_{n+1}^\beta \delta(E_n^\alpha-E_{n+1}^\beta-\Omega)\theta(E_n^\alpha)\theta(-E_{n+1}^\beta)\nonumber\\
&\quad +\sum_{\alpha,m}W_n^\alpha W_{m,n+1} \delta(E_n^\alpha-P_{m,n+1}-\Omega)\theta(E_n^\alpha)\theta(-P_{m,n+1})\nonumber\\
&\quad +\sum_{m,\beta}W_{m,n}W_{n+1}^\beta \delta(P_{m,n}-E_{n+1}^\beta-\Omega)\theta(P_{m,n})\theta(-E_{n+1}^\beta)\nonumber\\
&\quad +\sum_{m,m'}W_{m,n}W_{m',n+1} \delta(P_{m,n}-P_{m',n+1}-\Omega)\theta(P_{m,n})\theta(-P_{m',n+1})\bigg]\nonumber\\
&\quad +(1\pm1)[ n\to n+1, n+1\to n]  + (1\mp1)[ n\to -n, n+1\to -(n+1)] \nonumber\\
&\quad +(1\pm1)[ n\to -(n+1), n+1\to -n] + (1\mp1)[ n\to n, n+1\to -(n+1)] \nonumber\\
&\quad +(1\pm1)[ n \to -(n+1), n+1\to n] +(1\mp1)[ n\to -n, n+1\to n+1]\nonumber\\
&\quad +(1\pm1)[ n\to n+1, n+1\to -n]
\Bigg\rbrace,
\end{align}
where the terms such as $[ n\to n+1, n+1\to n]$ indicate the four terms in the first set of square brackets repeated with the indicated changes. The $\delta$ functions in this expression tell us which transitions contribute, and the Heaviside functions tell us when those transitions can occur (i.e., that the initial state must be below the Fermi energy and final state above). Many of the transitions are insignificant, however. We can neglect transitions between two phonon-assisted peaks, since they are suppressed by a factor of $W_{m_i,n_i}W_{m_f,n_f}\lesssim \lam^2/(1+\lam)^2$. We can also neglect ``inverted" transitions of the form $-n\to-(n+1)$, $n+1\to n$, $n\to -(n+1)$, and $n+1\to -n$. Such transitions, which go from a higher index to a lower one, are obviously impossible in the bare case, since the level with a higher index cannot fall below the Fermi energy while the one with a lower index falls above it. In the dressed case, such transitions become possible, but they have negligible weight, as discussed in Appendix~\ref{inverted_transitions}. Dropping these terms, we have
\begin{align}
\spm(\Omega) &=\sigma_0\frac{M_1^2}{\Omega}\sum_{n=0}^\infty\Bigg\lbrace
(1\pm1)\Bigg[\sum_{\alpha,\beta}W_{n+1}^\alpha W_{n}^\beta \delta(E_{n+1}^\alpha-E_{n}^\beta-\Omega)\theta(E_{n+1}^\alpha)\theta(-E_{n}^\beta)\nonumber\\
&\quad +\sum_{\alpha,m}W_{n+1}^\alpha W_{m,n} \delta(E_{n+1}^\alpha-P_{m,n}-\Omega)\theta(E_{n+1}^\alpha)\theta(-P_{m,n})\nonumber\\
&\quad +\sum_{m,\beta}W_{m,n+1}W_{n}^\beta \delta(P_{m,n+1}-E_{n}^\beta-\Omega)\theta(P_{m,n+1})\theta(-E_{n}^\beta)\Bigg]\nonumber\\
&\quad + (1\mp1)\Bigg[\sum_{\alpha,\beta}W_{-n}^\alpha W_{-(n+1)}^\beta \delta(E_{-n}^\alpha-E_{-(n+1)}^\beta-\Omega)\theta(E_{-n}^\alpha)\theta(-E_{-(n+1)}^\beta)\nonumber\\
&\quad +\sum_{\alpha,m}W_{-n}^\alpha W_{m,-(n+1)} \delta(E_{-n}^\alpha-P_{m,-(n+1)}-\Omega)\theta(E_{-n}^\alpha)\theta(-P_{m,-(n+1)})\nonumber\\
&\quad+\sum_{m,\beta}W_{m,-n}W_{-(n+1)}^\beta \delta(P_{m,-n}-E_{-(n+1)}^\beta-\Omega)\theta(P_{m,-n})\theta(-E_{-(n+1)}^\beta)\Bigg]\nonumber\\
&\quad+ (1\pm1)\Bigg[\sum_{\alpha,\beta}W_{n+1}^\alpha W_{-n}^\beta \delta(E_{n+1}^\alpha-E_{-n}^\beta-\Omega)\theta(E_{n+1}^\alpha)\theta(-E_{-n}^\beta)\nonumber\\
&\quad +\sum_{\alpha,m}W_{n+1}^\alpha W_{m,-n} \delta(E_{n+1}^\alpha-P_{m,-n}-\Omega)\theta(E_{n+1}^\alpha)\theta(-P_{m,-n})\nonumber\\
&\quad+\sum_{m,\beta}W_{m,n+1}W_{-n}^\beta \delta(P_{m,n+1}-E_{-n}^\beta-\Omega)\theta(P_{m,n+1})\theta(-E_{-n}^\beta)\Bigg]\nonumber\\
&\quad + (1\mp1)\Bigg[\sum_{\alpha,\beta}W_{n}^\alpha W_{-(n+1)}^\beta \delta(E_{n}^\alpha-E_{-(n+1)}^\beta-\Omega)\theta(E_{n}^\alpha)\theta(-E_{-(n+1)}^\beta)\nonumber\\
&\quad +\sum_{\alpha,m}W_{n}^\alpha W_{m,-(n+1)} \delta(E_{n}^\alpha-P_{m,-(n+1)}-\Omega)\theta(E_{n}^\alpha)\theta(-P_{m,-(n+1)})\nonumber\\
&\quad+\sum_{m,\beta}W_{m,n}W_{-(n+1)}^\beta \delta(P_{m,n}-E_{-(n+1)}^\beta-\Omega)\theta(P_{m,n})\theta(-E_{-(n+1)}^\beta)\Bigg]
\Bigg\rbrace,\label{sigma_general}
\end{align}
\end{widetext}
From this, one can obtain $\sxx$ by making the replacements $(1\pm1)\to 1$ and $(1\mp1)\to 1$, and $\sxy$ by making the replacements $(1\pm1)\to \mp 1$ and $(1\mp1)\to \pm 1$. One can obtain the result for the bare case by eliminating all sums over $\alpha$ and $\beta$, setting $W_n=1$ and $E_n=M_n-\mu_0$ for all $n$, and setting the weight of phonon-assisted peaks $W_{m,n}$ to zero for all $m$ and $n$.

Equation \eqref{sigma_general} represents eight possible types of transitions: intraband transitions within the conduction band, between either two quasiparticle levels or a quasiparticle level and a phonon-assisted peak,
\begin{align}
n^\beta\to (n+1)^\alpha, \label{transition1}\\
n_p^m\to (n+1)^\alpha, \label{transition2}\\
n^\beta\to (n+1)_p^m, \label{transition3}
\end{align}
which correspond to the first three lines of Eq.~\eqref{sigma_general}; intraband transitions within the valence band,
\begin{align}
[-(n+1)]^\beta\to -n^\alpha, \label{transition4}\\
[-(n+1)]_p^m\to -n^\alpha, \label{transition5}\\
[-(n+1)]^\beta\to -n_p^m, \label{transition6}
\end{align}
which correspond to the fourth, fifth, and sixth lines of Eq.~\eqref{sigma_general}; interband transitions of the form $-n\to n+1$,
\begin{align}
(-n)^\beta\to (n+1)^\alpha, \label{transition7}\\
(-n)_p^m\to (n+1)^\alpha, \label{transition8}\\
(-n)^\beta\to (n+1)_p^m, \label{transition9}
\end{align}
which correspond to the seventh, eighth, and ninth lines of Eq.~\eqref{sigma_general}; and interband transitions of the form $-(n+1)\to n$,
\begin{align}
[-(n+1)]^\beta\to n^\alpha, \label{transition10}\\
[-(n+1)]_p^m\to n^\alpha, \label{transition11}\\
[-(n+1)]^\beta\to n_p^m, \label{transition12}
\end{align}
which correspond to the tenth, eleventh, and twelfth lines of Eq.~\eqref{sigma_general}. Unless the level spacing at the Fermi energy is comparable to the phonon energy, the transitions that end at a phonon-assisted peak---\eqref{transition3}, \eqref{transition6}, \eqref{transition9}, and \eqref{transition12}---will be insignificant, since the initial level must fall below the Fermi energy and the phonon-assisted peak must fall at or above the phonon energy.


\section{Inverted transitions}\label{inverted_transitions}

\begin{figure}[tb]
\begin{center}
\includegraphics[width=\columnwidth]{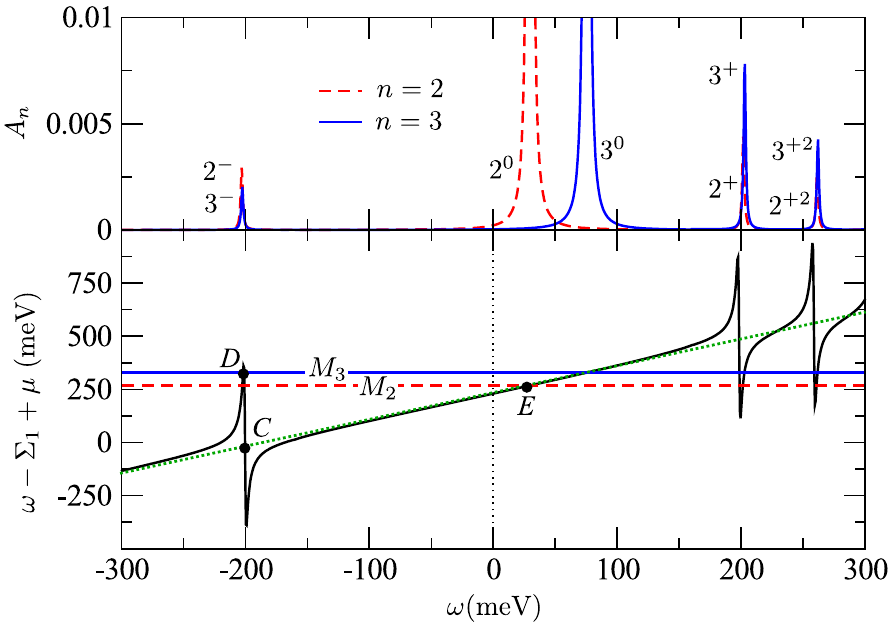}
\end{center}
\caption{(Color online) Splitting of levels such that inverted transitions are possible for  $B=27.4$T, $\mu_0=230$meV, $\Gamma=1.0$meV, $\omega_E=160$meV, and $A=300$meV. The upper frame shows $A_2(\omega)$ and $A_3(\omega)$, with the $3^-$ peak in $A_3$ falling below the Fermi energy and the $2^0$ peak (among others) falling above the Fermi energy, making a transition $3^-\to2^0$ possible. The lower frame displays $\omega-\Sigma_1+\mu$, showing how this set of peaks arise. The vertical dotted line marks the Fermi energy, and the dotted green line represents the $B=0$ behavior about which $\omega-\Sigma_1+\mu$ oscillates. The points C, D, and E mark points relevant to deriving the constraints on when an inverted transition is possible, as discussed in the text.\label{inverted}}
\end{figure}

By an ``inverted" transition, we mean a transition in which the index $n_f$ of the final state is less than the index $n_i$ of the initial state. Here we show that while such transitions are possible, due to splitting of levels, they have negligible optical weight. For simplicity, we consider the conditions under which an intraband inverted transition is possible; interband inverted transitions will obviously have at least as stringent conditions. We also consider only transitions between quasiparticle levels, not between a level and a phonon-assisted peak. Figure~\ref{inverted} illustrates a situation allowing an inverted transition of the form $3\to2$. The top frame shows the spectral functions $A_2$ and $A_3$, and we see the quasiparticle level $3^-$ lies below the Fermi energy while the peak $2^0$ lies above. According to the selection rules derived in Appendix.~\ref{selection_rules}, this means the transition $3^-\to 2^0$ is allowed. However, the $3^-$ level has negligible weight, meaning the transition will as well. The transition $3^-\to2^+$, which we can see is also possible, contains even less weight.

This low weight is generic: the conditions allowing the transitions are so extreme that they preclude two strong peaks between which the transition can occur. Examine the lower frame of Fig.~\ref{inverted} and consider a transition $n+1\to n$ rather than the specific case $3\to2$ shown. In order for the inverted transition to occur, an intersection of $M_{n+1}$ with $\omega-\Sigma_1+\mu$ [marking a level $(n+1)^\beta$] must occur to the left of the Fermi energy (marked by a vertical dotted line), and an intersection of $M_{n}$ with $\omega-\Sigma_1+\mu$ (marking a level $n^\alpha$) must occur to the right of the Fermi energy. We can easily see from the figure that if this occurs, the peak $(n+1)^\beta$ will be separated from the principle peak in $A_{n+1}$---the peak at roughly $(M_{n+1}-\mu_0)/(1+\lam)$---by at least an energy $\omega_E$. Since the weight in a peak decays with the peak's separation from $(M_{n+1}-\mu_0)/(1+\lam)$, this peak will always have negligible weight. The reader should be able to convince him or herself that any inverted transitions would rely on such a large splitting and therefore be insignificant.

For completeness, we derive explicitly the conditions under which an inverted transition such as that in Fig.~\ref{inverted} can occur. Referring to the lower frame, the point D labels the intersection of $\omega-\Sigma_1+\mu$ with $M_{n+1}$ (in this case, $M_3$); E, the intersection with $M_n$ (in this case, $M_2$). The point C labels the midpoint of an oscillation in $\Sigma$, occuring at an energy $P_m$. For E to lie to the right of $E_F$, we need $E_n^\alpha>0$. Taking this energy to fall below the phonon energy, we can use $E_n^\alpha=(M_n-\mu_0)/(1+\lam)$ and obtain the constraint
\begin{equation}
\mu_0<M_n.\label{constraint1}
\end{equation}
For D to lie to the left of $E_F$, we need $M_{n+1}$ to lie below the peak height of the oscillation at $P_m$. This height can be calculated as the vertical position of C plus the amplitude of the oscillation. The vertical position of C is approximately $(1+\lam)P_m+\mu_0$, given that $\omega-\Sigma_1+\mu$ can be approximated by the green line, which is obtained by letting $\Sigma_1(\omega)=\Sigma_1(0)-\lam\omega$. The amplitude of the oscillation is given by\cite{Pound:11b} $\frac{AM_1^2}{4W_C\Gamma}$. So the condition that $M_{n+1}$ falls below the peak height of the oscillation becomes $M_{n+1}<(1+\lam)P_m+\mu_0+\frac{AM_1^2}{4W_C\Gamma}$. Rearranging this, we obtain a constraint on the broadening $\Gamma$,
\begin{equation}
\Gamma<\frac{AM_1^2}{4W_C}\left[M_{n+1}-(1+\lam)P_m-\mu_0\right]^{-1}.\label{constraint2}
\end{equation}
In addition to Eqs.~\eqref{constraint1} and \eqref{constraint2}, there are constraints obtained from the location of the oscillation in $\Sigma_1$: $P_m$ must be less than $-\omega_E$, and we wish there to be no oscillation that occurs between $P_m$ and  $-\omega_E$. Using $P_m=-\omega_E+M_m-\mu_0$, we obtain
\begin{equation}
M_m<\mu_0<M_{m+1}.\label{constraint3}
\end{equation}
The parameters in Fig.~\ref{inverted} were found by fixing $n$ and all other parameters except $\Gamma$ and $\mu_0$, making $m$ a function of $\mu_0$ via Eq.~\eqref{constraint3}, and then varying $\mu_0$ to maximize the value of $\Gamma$ allowed by Eq.~\eqref{constraint2}. Even with that optimization, $\Gamma$ must be very small in order to allow an inverted transition. 

Although we have only dealt explicitly with the level-level inverted transitions, the same form of argument shows that analogous transitions involving phonon-assisted peaks are also insignificant.

\bibliography{optics}
\end{document}